\documentclass[a4paper,11pt]{article}
\usepackage{jinstpub} 
\usepackage{lineno}
\usepackage[version=4]{mhchem}
\usepackage{graphicx}
\usepackage{microtype}
\usepackage{xurl}
\usepackage{subcaption}
\usepackage{lipsum}
\usepackage{amsmath}
\usepackage{ulem}
\usepackage{float}
\usepackage{caption}
\usepackage{booktabs}
\usepackage{siunitx}
\usepackage{tikz}
\usepackage{booktabs}
\usepackage{pgfplots}
\usepackage{hyperref}

\title{\boldmath A High Gain Preamplifier Board for Low Charge Semiconductor Detectors}

\author[a]{D. Zhang,}
\author[a]{K. Ma,}
\author[a]{J. Zhang,}
\author[b]{J. Ge,}
\author[a]{Y. Liu,}
\author[a,1]{H. Liang\note{Corresponding author.}}
\emailAdd{simonlh@ustc.edu.cn}
\affiliation[a]{{Department of Modern Physics}, {University of Science and Technology of China},
{ Hefei},{ 230026}, { China}}
\affiliation[b]{{Department of Physics}, {University of Michigan},
{ Ann Arbor},{ MI 48109}, { USA}}

\abstract{We present a high-gain, low-noise preamplifier board designed for reading out low-charge semiconductor detectors—specifically Low-Gain Avalanche Detectors (LGADs) and three-dimensional (3D) silicon sensors—in high-energy physics applications. The circuit employs a three-stage architecture featuring a discrete SiGe:C bipolar junction transistor (BJT)-based transimpedance amplifier (TIA) front-end followed by two resistive feedback amplification stages using LTC6431 chips. This configuration achieves a charge gain of \qty{116.31}{\milli\volt  \cdot \nano\second \per \femto\coulomb} with excellent linearity over an input range of \SIrange[]{0.5}{20}{\femto\coulomb} and a wide bandwidth spanning from 34.0 MHz to 594.3 MHz. Experimental evaluations coupled with LGAD detectors demonstrate a time resolution of 34.93 ps and an equivalent noise charge (ENC) of 0.19 fC at 20 °C. Furthermore, tests with conventional PIN sensors without a gain layer yield a timing resolution of 77.78 ps with an ENC of 0.11 fC, and tests with the 3D silicon detector yield 38.66 ps with an ENC of 0.09 fC, all measured under the same thermal conditions. All configurations confirm the board's capability in low-signal regimes, significantly showing improved performance. A six-channel variant of the board has also been developed to support position-sensitive measurements. These findings demonstrate the board’s suitability for laboratory-based 4D tracking detector characterisation, while simultaneously providing the groundwork for dedicated ASIC development.}


\begin{document}
\maketitle
\flushbottom

\section{Introduction}
\label{sec:intro}

High-energy physics experiments at the Large Hadron Collider (LHC) and its future upgrades demand unprecedented precision in particle tracking to explore fundamental physics beyond the Standard Model~\cite{a}. Detector systems must simultaneously satisfy stringent requirements: sub-100 ps timing resolution, micron-level spatial resolution, exceptional radiation hardness, high detection efficiency, and long-term operational stability~\cite{b}. These demands are particularly acute for the High-Luminosity LHC (HL-LHC) phase, where the pile-up event rate will increase by an order of magnitude compared to the current LHC operation~\cite{Sicking:2019}.

To meet these challenges, novel semiconductor detector technologies have been developed. The LGAD, proposed by the RD50 collaboration, provides a promising solution by providing internal gain (typically 10--30) while maintaining excellent time resolution ($\sim$40 ps)~\cite{c}. While DC-LGAD achieves superior timing performance, its spatial resolution is limited by the pad size (typically $\sim$ 1 mm). In contrast, AC-LGAD, by employing a resistive charge-sharing readout scheme, can determine the hit position through signal interpolation among neighboring electrodes, thereby achieving a spatial resolution at the level of  \qtyrange{10}{50}{\micro\metre} while maintaining excellent timing performance ($\sim$ \qty{30}{\pico\second})~\cite{li_2026}.

Complementing LGAD technology, 3D silicon detectors embed electrodes directly into the silicon bulk, decoupling the electrode distance from the substrate thickness~\cite{f,g}. Achieving a sufficiently short electrode distance through the layout design offers high spatial resolution, high radiation tolerance due to the mitigation of charge trapping effects~\cite{davia2009319} and fast signal response without relying on a gain layer. However, 3D detectors generate very small charge outputs, approximately several fC or even below 1 fC, respectively placing extreme demands on the front-end electronics. Furthermore, their rapid rise times (potentially < 500 ps) require front-end amplifiers with exceptionally high bandwidth ($\sim$ 1 GHz) to avoid signal distortion and preserve timing resolution~\cite{h}.

Recognizing these challenges, the community has devoted considerable effort to developing specialized front-end electronics. In the ASIC domain, the ALPIDE chip~\cite{9508095} for the ALICE ITS upgrade, the ALTIROC~\cite{agapopoulou_2020} for the ATLAS HGTD, the ETROC~\cite{etroc} for the CMS MTD ETL, the EICROC~\cite{eicroc} for the future EIC, and the Timepix family~\cite{timepix} for hybrid pixel detectors have all achieved notable success in low-noise and high-speed processing. Discrete solutions, such as the UCSC preamplifier board and the FNAL preamplifier board, represent the most widely used approach for LGAD readout. Despite these advances, a common limitation persists: most existing designs are tailored to a single detector type, and even the most sensitive discrete solutions have yet to demonstrate reliable performance for charge outputs below 1 fC—a critical requirement for 3D detectors with small signal. This gap motivates our work.

This paper presents the design and characterization of a novel front-end amplifier board specifically engineered to address these challenges. Our board features a low-noise, high-bandwidth preamplifier stage tailored for signals from both LGAD and 3D silicon detectors. We report comprehensive performance evaluations, including noise analysis, linearity tests, and timing resolution measurements. The results demonstrate that our amplifier achieves the necessary signal-to-noise ratio (SNR) and bandwidth, paving the way for high-performance 4D tracking applications.
\section{ Electronic design}
To simultaneously achieve higher bandwidth and gain, it is evident that single-stage amplification alone cannot satisfy the system requirements. Consequently, the implementation of multiple amplification stages becomes imperative to enhance the overall system gain. Additionally, since detectors typically generate current signals~\cite{1686997}, the conversion of these current signals into voltage signals constitutes a critical processing step.

\subsection{Amplifier architecture}
Charge-sensitive amplifier(CSA) and transimpedance amplifier(TIA) represent the predominant circuit configurations for this application, each demonstrating distinct operational advantages. The charge-sensitive amplifier fundamentally employs an integrating feedback capacitor as its core component. This configuration exhibits significant insensitivity to the sensor's inherent capacitance and demonstrates superior performance in suppressing current noise, particularly flicker noise and current-related components of amplifier noise~\cite{i}. However, bandwidth is limited by the feedback capacitor discharge mechanism, thereby imposing limitations on transient response characteristics. In contrast, the transimpedance amplifier utilizes a feedback resistor as its fundamental element. This architecture enables the attainment of bandwidths extending into the GHz range through judicious selection and optimization of the amplifier. Of particular significance is its ability to accurately reproduce the input pulse shape at the output, thereby maintaining temporal fidelity without introducing signal distortion.

Based on these considerations, taking into account the specific output characteristics of our high-precision timing detector and the critical requirements for wide bandwidth and pulse shape preservation, the TIA emerges as the optimal choice for the front-end stage.

\begin{figure*}[ht]
	\centering
	\includegraphics[width=0.8\textwidth]{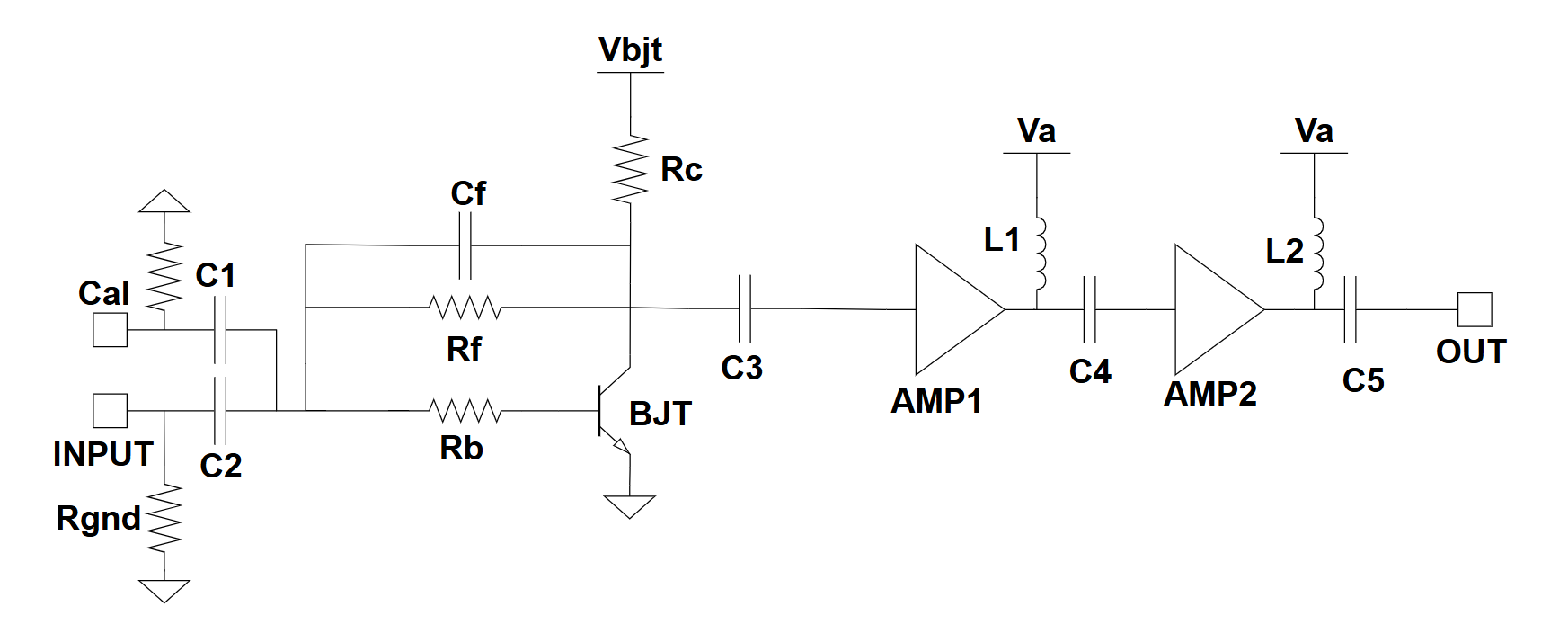}
	\caption{Simplified schematic of the preamplifier.}
	\label{fig:schematic-label}
\end{figure*}

\autoref{fig:schematic-label} illustrates the simplified schematic of our three-stage amplification circuit. Generally, the impact of an amplifier stage on overall system performance is inversely proportional to its distance from the detector; hence, the first stage, being the closest, demands the most meticulous component selection. For the TIA, the bandwidth is primarily determined by the following critical factors~\cite{TI_ssztbc4}:
\begin{equation}
	f_{-3\mathrm{dB}} = \sqrt{\frac{\mathrm {GBWP}}{2\uppi R_f C_{\text{in}}}},
\end{equation}
where $\text{GBWP}$ is Gain Bandwidth Product, $C_{\text{in}}$ is the total input capacitance.This input capacitance comprises three primary components: the detector capacitance, PCB parasitic capacitance, and the input capacitance of the first-stage active device. These parameters can be considered fixed upon completion of the circuit design.

To achieve higher bandwidth, an active device with an adequate $\text{GBWP}$ is essential. Among commercially available  amplifiers, the OPA855 currently offers the highest GBWP of 8 GHz for small-signal output. For a target bandwidth of approximately 1 GHz with an input capacitance of  $\sim$10 pF, the required feedback resistor  ($R_f$) would need to be approximately 120 $\Omega$. However, this value is often impractical for most applications. Consequently, rather than utilizing commercial off-the-shelf amplifiers, we employed a discrete SiGe:C NPN BJT, specifically the BFP840FESD, which features a transition frequency of 85 GHz~\cite{Infineon_BFP840FESD}. The base--collector and collector--emitter parasitic capacitances of the BFP840FESD are $C_{cb} = 38$~fF and $C_{ce} = 0.37$~pF, respectively. This selection enabled us to implement a transimpedance gain resistor of 1.3~k$\Omega$, more effectively satisfying our amplification requirements. A feedback capacitor $C_f$ of approximately 120~fF is required in parallel with $R_f$, setting the closed-loop $-3$~dB bandwidth to

\begin{equation}
f_{-3\mathrm{dB}} \approx \frac{1}{2\pi R_f C_f} \approx 1~\text{GHz}.
\end{equation}

In practice, this capacitance can be fully provided by the parasitic capacitance inherent to the PCB layout and the BJT collector node, eliminating
the need for an additional discrete component. It should be noted that the selection of the first-stage device is not exclusively determined by bandwidth considerations. Practical implementation also necessitates careful evaluation of noise performance, nonlinearity, and offset characteristics. The BFP840FESD was ultimately selected as the core component in our design due to its superior overall performance across these critical parameters.

Following current-to-voltage conversion in the first stage, the primary objectives of subsequent amplification stages are to: amplify the voltage signal to an appropriate level, maintain optimal SNR, and  minimize voltage distortion and offset. The overall bandwidth of the preamplifier system is governed by the cascaded bandwidth characteristics of each stage, with the narrowest bandwidth stage acting as the limiting factor. After comprehensive evaluation, the LTC6431 was selected for its superior performance characteristics, including a power gain of 20.8 dB, 2 GHz of $f_{\mathrm{-3dB}}$, and excellent noise performance. The LTC6431 implementation requires careful consideration of several design aspects:
\begin{itemize}
	\item DC-blocking capacitors to isolate external bias interference.
	\item Power supply inductors to mitigate high-frequency noise coupling from DC sources.
\end{itemize}

Furthermore, the inherent LC network introduces frequency-dependent filtering effects, necessitating precise optimization of capacitor and inductor values based on the target operational frequency range.

\subsection{More considerations}

\subsubsection{Noise}

In addition to the judicious selection of devices for noise reduction as previously discussed, a comprehensive approach is required to effectively mitigate noise interference. Beyond intrinsic electronic noise, device performance is significantly influenced by DC power supply characteristics. Notably, excessive power supply noise can detrimentally impact signal integrity in the circuit. Particular attention must be devoted to the power supply for first-stage transistors, as these components exhibit heightened noise sensitivity and limited power supply rejection capability. The LT3042 power regulator was selected for this stage, featuring an impressive RMS noise of $\SI{0.8}{\micro\volt}$ and a power supply rejection ratio (PSRR) of \SI{79}{\deci\bel}. For subsequent stages (second and third), the ADM7171 regulator was implemented, demonstrating an RMS noise performance of approximately $\SI{5}{\micro\volt}$.

Another crucial factor to consider involves the potential noise interference from external environmental sources, such as 5G signals emitted by mobile devices and electromagnetic noise generated by large electronic appliances. Given the impracticality of maintaining the testing environment within a fully shielded metal enclosure at all times, a compact shielding enclosure was specifically designed and implemented above the electronic components to significantly mitigate external environmental interference.

\subsubsection{Calibration}

To accurately determine the charge output corresponding to the signal amplitude from the detector, merely converting and amplifying the signal into a voltage signal is insufficient. Additionally, the performance variation of the preamplifier across different temperatures necessitates calibration. Consequently, a dedicated injection current channel is implemented adjacent to the input port. By injecting a fast voltage pulse with a well‑defined amplitude, a corresponding charge injection is generated across capacitor $C_1$. This circuit configuration enables the calculation of the overall charge‑to‑voltage gain. However, to faithfully mimic the fast current pulse from the detector, the injected pulse must possess a sufficiently short rise time; otherwise, the measured gain may deviate from the true response to detector signals. Accordingly, the current implementation still exhibits some dependence on the pulse shape, which highlights the need for careful pulse optimisation in future iterations.

\subsubsection{PCB layout}

\begin{figure}[ht]
	\centering
	\includegraphics[width=0.5\linewidth]{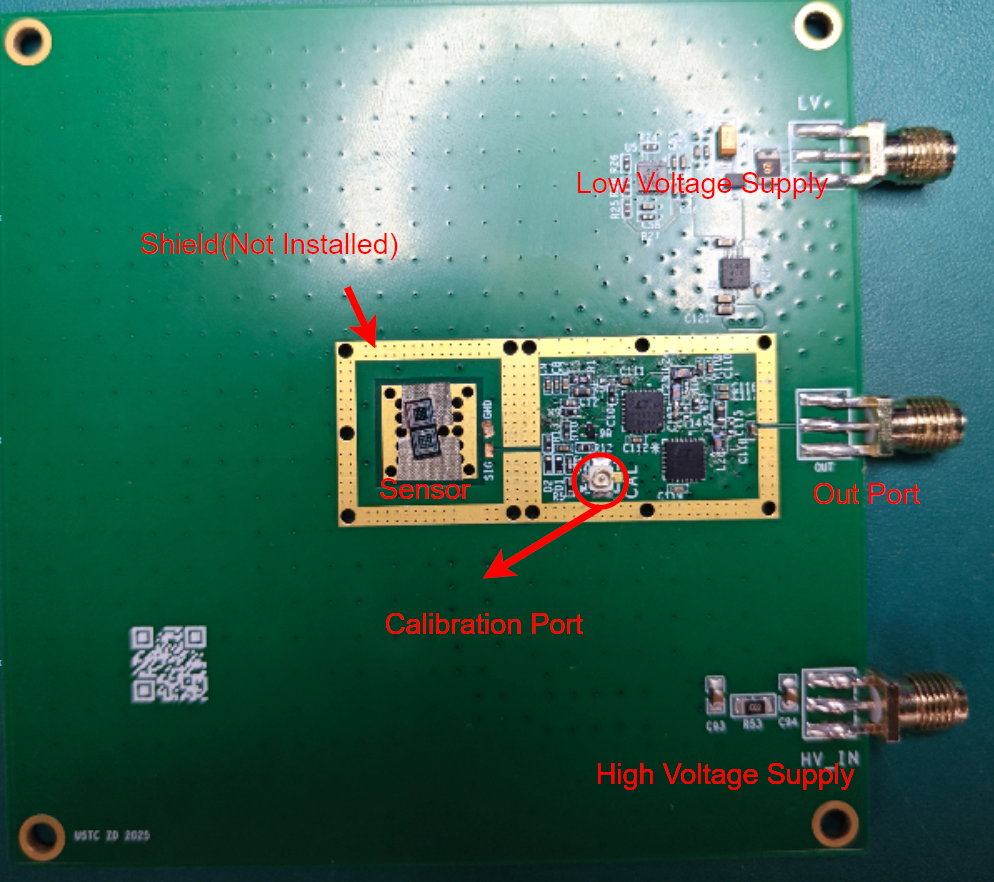}
	\caption{Photography of single-channel preamplifier.}
	\label{fig:single_channel_preamp-label}
\end{figure}

An optimal PCB layout design is essential for ensuring high signal integrity. In the first amplifying stage, a transimpedance amplifier (TIA) is employed
to convert the detector current into a voltage signal. Since the TIA input presents a virtual ground, conventional $50\,\Omega$ impedance matching is not applicable; the layout is instead optimized to minimize parasitic capacitance at the inverting input and to provide a low-impedance ground return path. Controlled-impedance routing and proper termination are adopted at the inter-stage connections to suppress signal reflections.  \autoref{fig:single_channel_preamp-label} shows the actual single-channel circuit board.

\subsection{Multi-channel preamplifier board}

\begin{figure}[ht]
	\centering
	\includegraphics[width=0.5\linewidth]{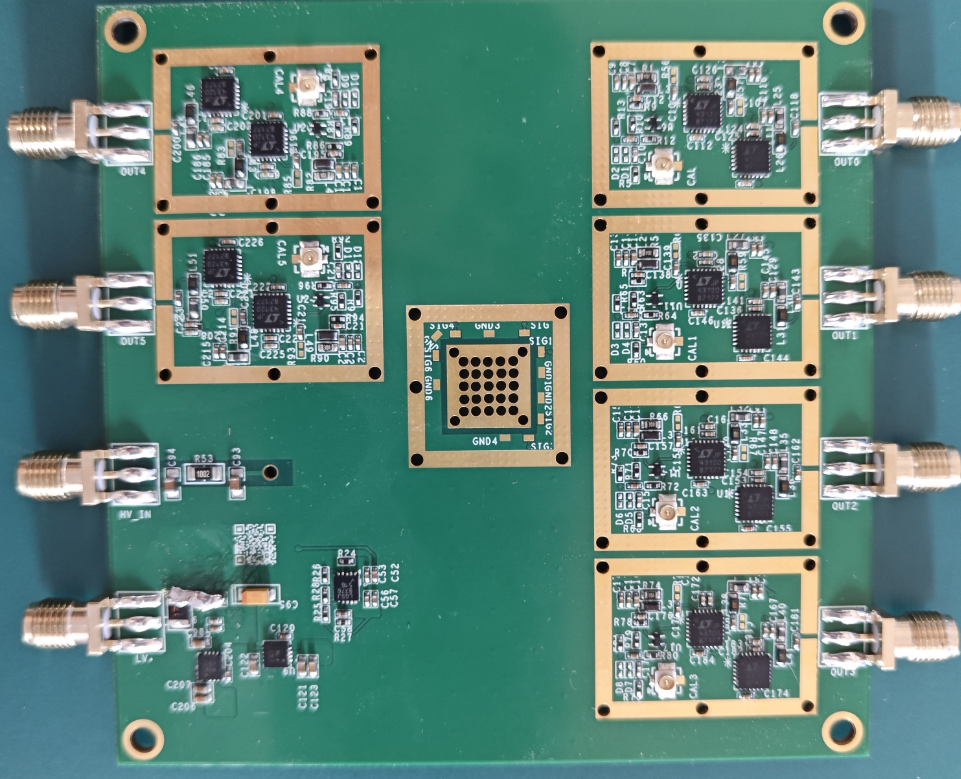}
	\caption{Photography of 6-channel preamplifier.}
	\label{fig:multi_channel_preamp-label}
\end{figure}

For applications requiring position resolution, a detector array comprising multiple pads is essential, wherein data from multiple channels must be processed to determine the particle's position through fitting algorithms~\cite{ott2023167541}. Additionally, temporal resolution can be improved by employing multiple channels in conjunction with position-time reconstruction techniques. Consequently, based on the single-channel preamplifier design, a multi-channel board was developed with an identical circuit architecture, achieving a 6-channel configuration as illustrated in  \autoref{fig:multi_channel_preamp-label}. This design enables multi-pad readout functionality. It should be noted that the increased channel count necessitates higher supply current; therefore, minor modifications were implemented in the DC low-voltage power supply circuit to accommodate this requirement.

\section{Test Results}

\subsection{Small-signal AC parameter test}

Prior to performing joint testing with the sensor, it is essential to evaluate the performance characteristics of the amplification board independently.

\subsubsection{Charge gain and linearity}

The initial phase involves comprehensive testing of charge gain, linearity, and dynamic range. For this purpose, an arbitrary waveform generator (WX2182C) is employed to generate high-speed pulse signals  (typical rise/fall time < 500 ps), which are injected through the reserved calibration port. The injected charge quantity can be subsequently calculated based on:

\begin{equation}
	Q_{\text{in}} = V_{\text{pulse}}*C_1,
\end{equation}

$V_{\text{pulse}}$ denotes the amplitude of the input pulse, while $C_1$ represents a 1 pF ceramic capacitor. The minimum pulse amplitude achievable by the signal generator is 50 mV, consequently yielding a minimum injectable charge of 50 fC. However, this value exhibits significant discrepancy with the actual charge generated by the sensor. To address this issue, a \SI{-40}{\deci\bel} attenuator was employed to reduce the signal voltage amplitude by two orders of magnitude, thereby achieving a minimum charge injection of  \qty{0.5}{\femto\coulomb}.

\begin{figure}[ht]
	\centering
	\includegraphics[width=0.6\linewidth]{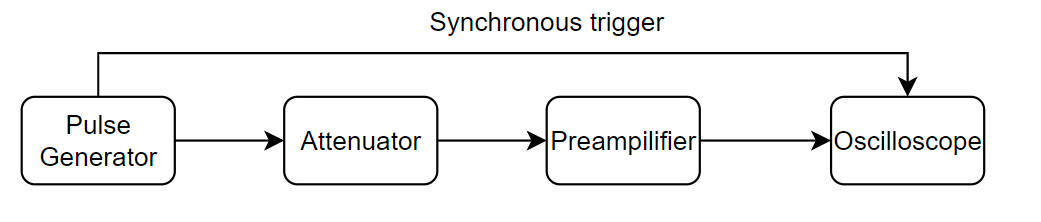}
	\caption{Setup of the charge gain and linearity test.}
	\label{fig:charge_pluse_integral-label}
\end{figure}

\autoref{fig:charge_pluse_integral-label} shows the complete experimental configuration. The output signal is acquired and recorded utilizing a LeCroy HDO9404 oscilloscope (featuring 4 GHz bandwidth and 40 GS/s sampling rate in 2-channel operation mode). To mitigate noise interference, 1000 waveform samples were collected at each voltage point, followed by signal acquisition at various output amplitude levels across different input voltage.

\begin{figure}[ht]
	\centering
	\includegraphics[width=0.5\linewidth]{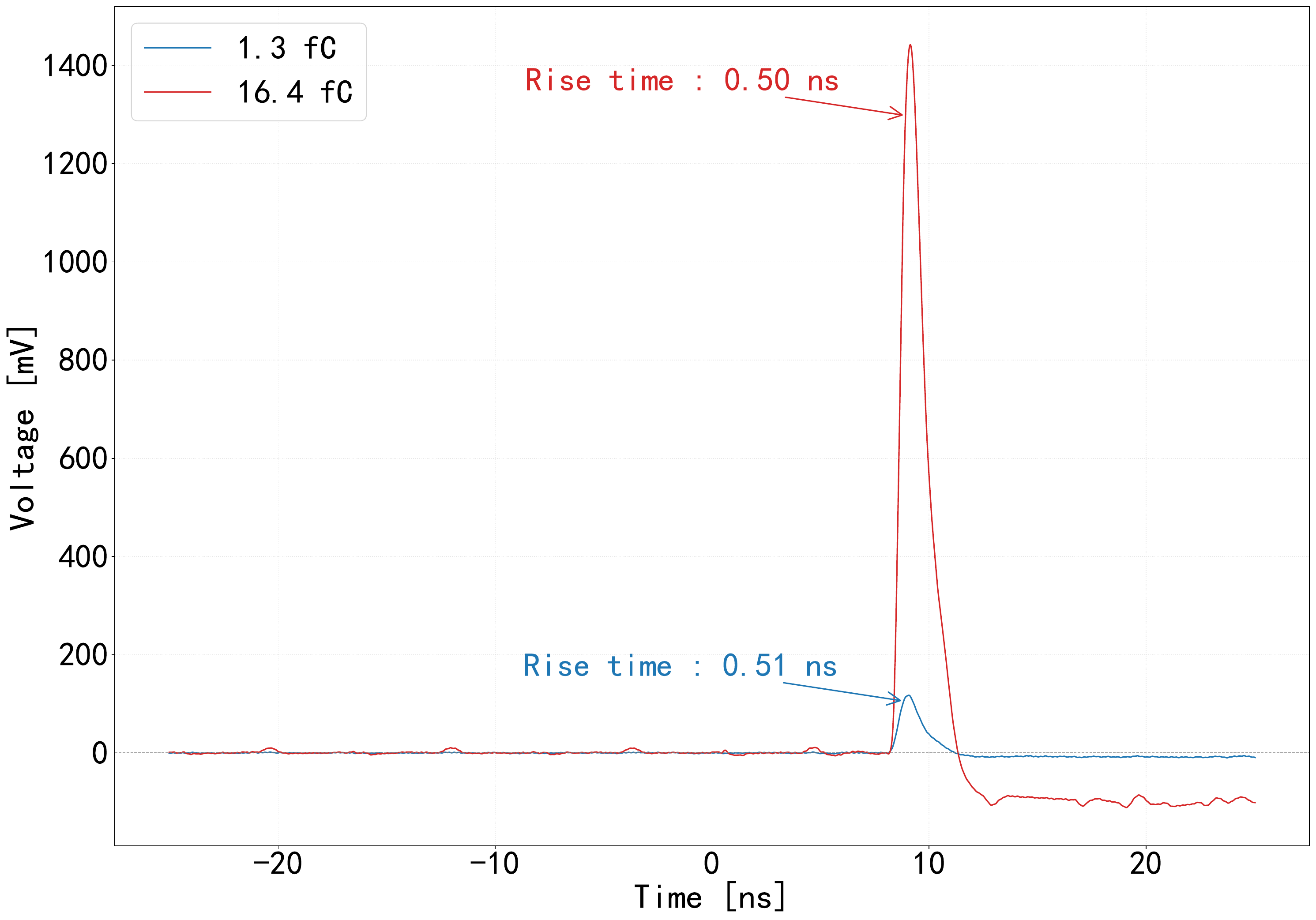}
	\caption{Typical pulse responses for two different charge injections}
	\label{fig:osci-label}
\end{figure}

To evaluate the dynamic range and the pulse shape consistency of the readout electronics, two representative intermediate charge injections, \SI{1.3}{\femto\coulomb} and \SI{16.4}{\femto\coulomb}, are selected for comparison. As shown in ~\autoref{fig:osci-label}, although the signal amplitudes differ by more than an order of magnitude, the pulse shapes remain similar, and the rise times, defined as the time for the signal to increase from 10\% to 90\% of its peak amplitude, remain consistent at approximately \SI{0.50}{\nano\second}. This indicates that the shaping bandwidth of the readout channel is highly stable and does not depend on the input charge within the measured dynamic range.

Prior to integration, a per-event pedestal baseline is subtracted using a 3$\sigma$ -clipped median of the pre-pulse region, with the noise level estimated via the median absolute deviation (MAD). The pulse area is then computed by trapezoidal integration over an adaptive window whose boundaries are set at a fraction of the peak amplitude, balancing full energy capture against noise inclusion. A tail-extension parameter is introduced to recover slow trailing components and is optimized by requiring stability of the extracted gain slope.

To ensure that the pulse integral faithfully represents the deposited charge, the integration window is restricted to the positive-pulse region and terminated before the onset of the undershoot. Specifically, the upper boundary is determined by locating the point at which the differentiated waveform first crosses zero after the pulse peak, marking the transition from the fast discharge to the slow baseline-recovery phase. No additional tail extension is applied beyond this adaptive boundary. To verify that the extracted charge gain is insensitive to the exact choice of integration limits, the upper boundary is systematically varied by $ \pm$0.8~ns around the nominal value. The resulting gain variation remains within 1.7\%, confirming the robustness of the chosen integration scheme.

\begin{figure}[ht]
	\centering
	\includegraphics[width=0.7\linewidth]{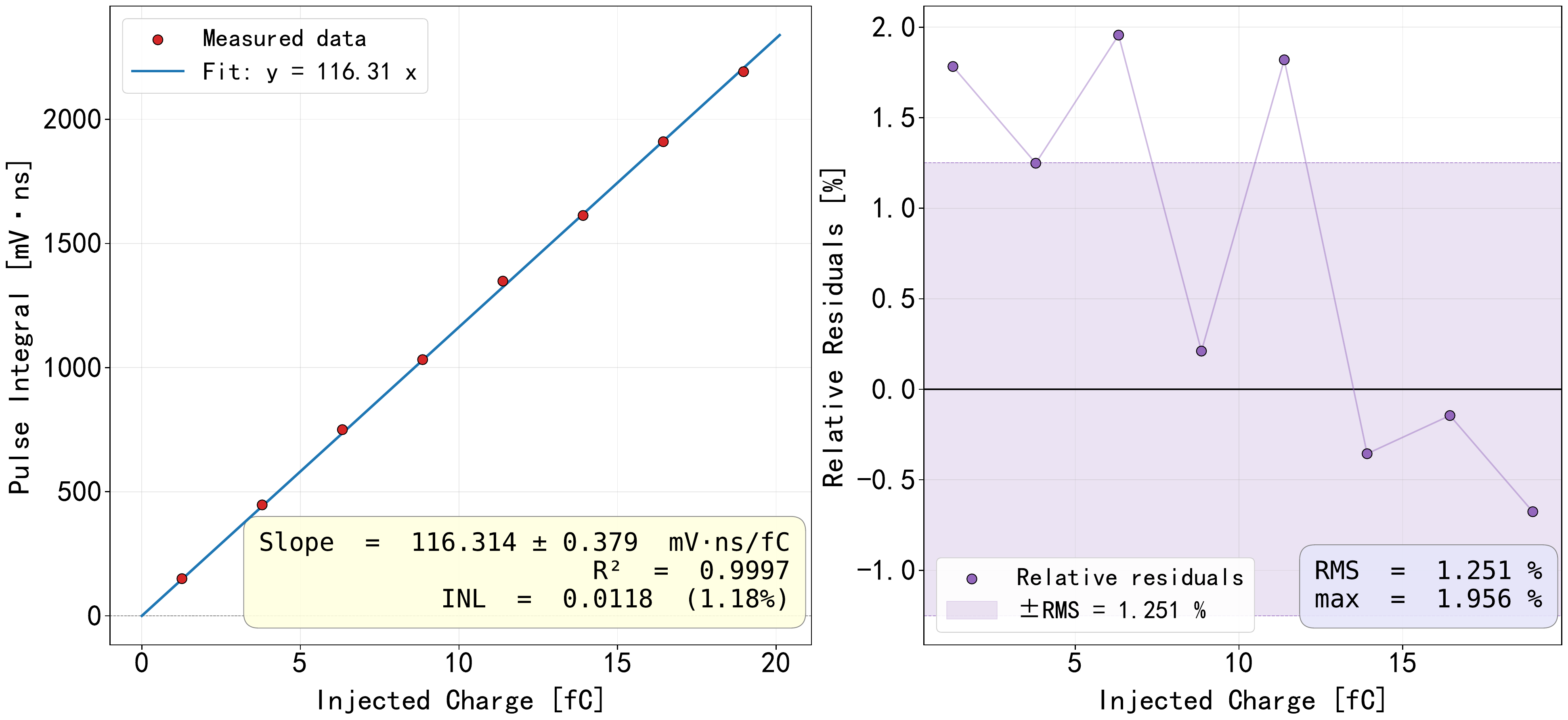}
	\caption{Integration of the averaged output waveform versus equivalent injected charge and fit residuals.}
	\label{fig:result_of_charge_gain-label}
\end{figure}

As shown in  \autoref{fig:result_of_charge_gain-label}, the test results yield a charge gain of \qty{116.31}{\milli\volt \cdot \nano\second \per \femto\coulomb}, with the measured charge range spanning from 0.50 fC to 20 fC. The $R^2$ and the maximum Integral Nonlinearity (INL) of the linear fit are 0.9997 and 1.18\% , respectively, which shows a good linearity.

\subsubsection{Frequency response}
\begin{figure}[ht]
	\centering
	\includegraphics[width=0.4\linewidth]{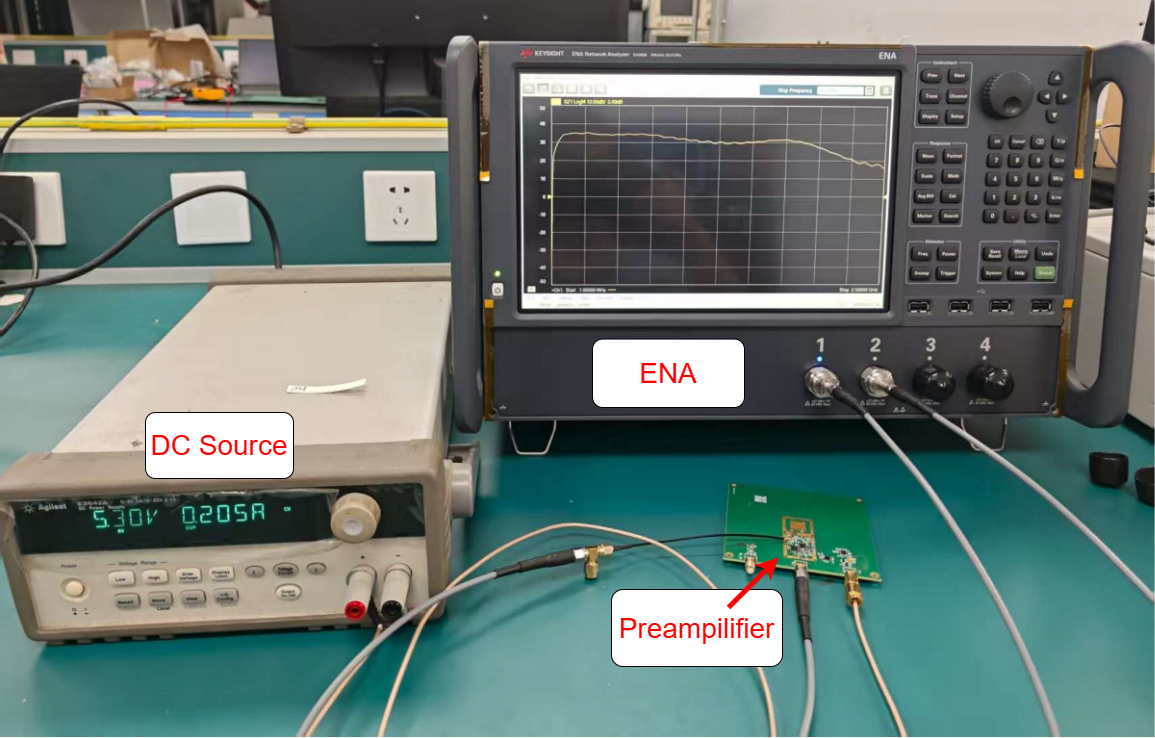}
	\caption{Setup for measuring frequency response.}
	\label{fig:setup_of_network_ana-label}
\end{figure}

To ensure proper operation of the sensor on the  preamplifier board, it is essential to characterize the frequency response across the entire operational bandwidth, thereby verifying that the target signal frequency falls within the passband. As illustrated in \autoref{fig:setup_of_network_ana-label}, frequency response analysis was conducted using a KEYSIGHT E5080B network analyzer (100 kHz - 26.5 GHz).
\begin{figure}[h]
	\centering
	\includegraphics[width=0.5\linewidth]{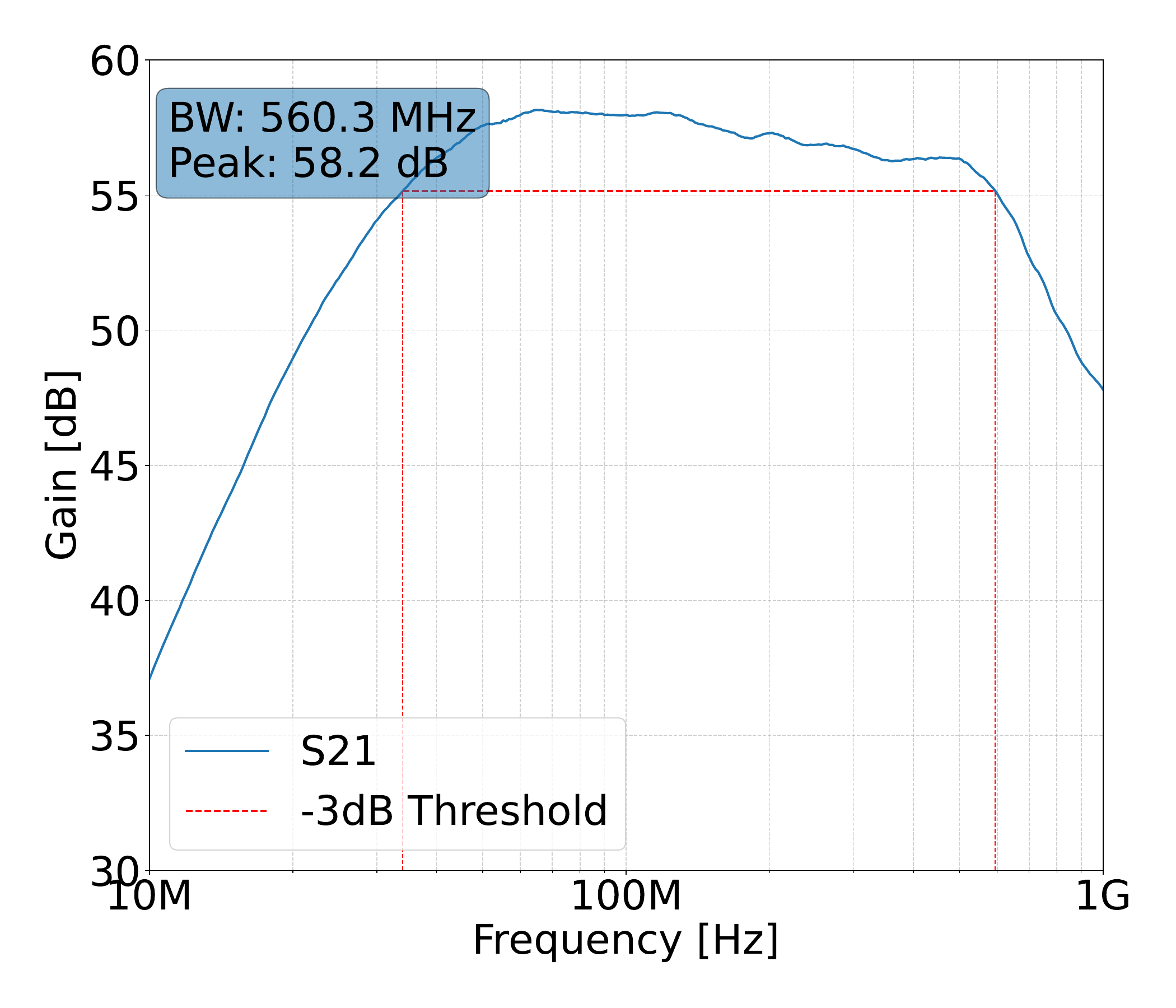}
	\caption{Frequency response curve of the single-channel preamplifier.}
	\label{fig:gain_freq_response-label}
\end{figure}

The measurement results are shown in \autoref{fig:gain_freq_response-label}. The measured frequency characteristics demonstrate a -3 dB bandwidth from 34.0 MHz to 594.3 MHz. The bandwidth was intentionally constrained to a level sufficient for the fast rising edge signals from LGAD and 3D silicon detectors, prioritizing an adequate SNR over achieving the maximum possible bandwidth.

\begin{figure}[h]
	\centering
	\includegraphics[width=0.5\linewidth]{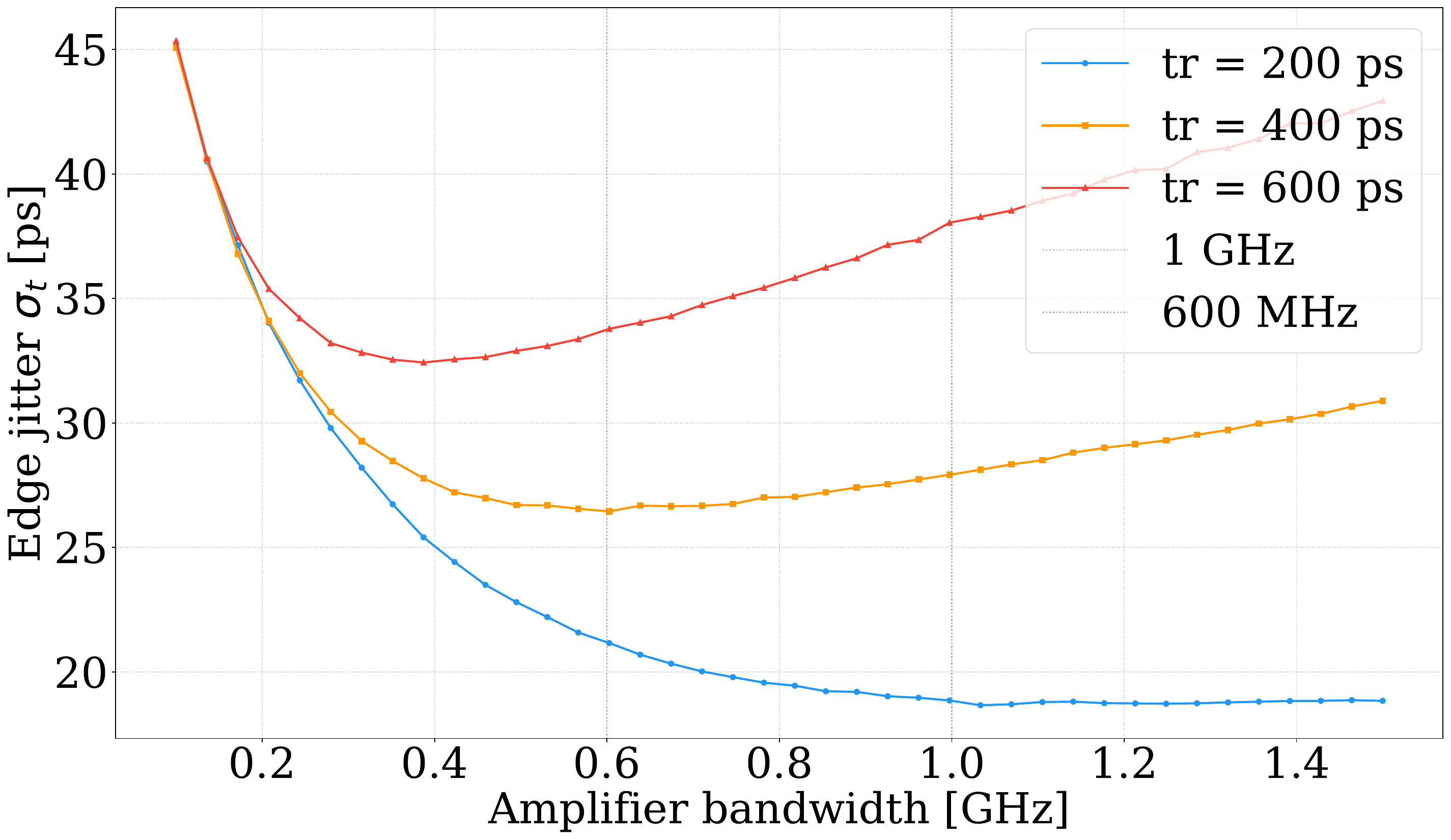}
	\caption{Effect of amplifier bandwidth and input signal rise time ($t_r$) on the rising edge timing jitter ($\sigma_t$)}
	\label{fig:jitter_bandwidth-label}
\end{figure}

Fig.~\ref{fig:jitter_bandwidth-label} illustrates the effect of amplifier bandwidth and source signal rise time ($t_r$) on the rising edge jitter ($\sigma_t$), assuming a fixed amplifier gain. The noise is modeled as uniformly distributed white noise, and the overall signal-to-noise ratio (SNR) is maintained at approximately 20 across different bandwidth settings, with minor variations arising from the bandwidth-dependent noise integration. The results are obtained from Monte Carlo simulations with $10^5$ trials for each data point, ensuring statistical convergence of the jitter distribution.

As shown in the figure, the edge jitter exhibits a characteristic U-shaped (bathtub) curve as a function of bandwidth for any given rise time. This behavior highlights the fundamental trade-off in bandwidth selection:

\begin{itemize}
    \item \textbf{Low-bandwidth regime:} When the amplifier bandwidth is limited, high-frequency components of the signal are severely attenuated. This stretches the rising edge and reduces its slope ($dV/dt$). According to the timing uncertainty relation $\sigma_t \propto \sigma_n / (dV/dt)$, a smaller slope directly amplifies the timing jitter, causing it to increase rapidly.

    \item \textbf{High-bandwidth regime:} Beyond a certain threshold, further increasing the bandwidth no longer improves the signal slope. Instead, it introduces more wideband noise. Since the total noise power is proportional to the bandwidth, the integrated noise increases, leading to a gradual rise in jitter.

    \item \textbf{Optimal bandwidth:} Each curve has a distinct minimum jitter point, representing the optimal bandwidth for a specific rise time. Notably, as $t_r$ increases (from 200 ps to 600 ps), the optimal bandwidth shifts to lower frequencies, and the minimum achievable jitter increases.
\end{itemize}

In conclusion, a larger amplifier bandwidth is not always better. Over-sizing the bandwidth degrades the timing resolution due to noise integration. The bandwidth should be carefully matched to the signal's rise time to achieve the best balance between signal fidelity and noise suppression.

\subsection{\texorpdfstring{$\beta$}--scope tests}
\subsubsection{Test results with LGAD}

\begin{figure}[ht]
	\centering
	\includegraphics[width=0.7\linewidth]{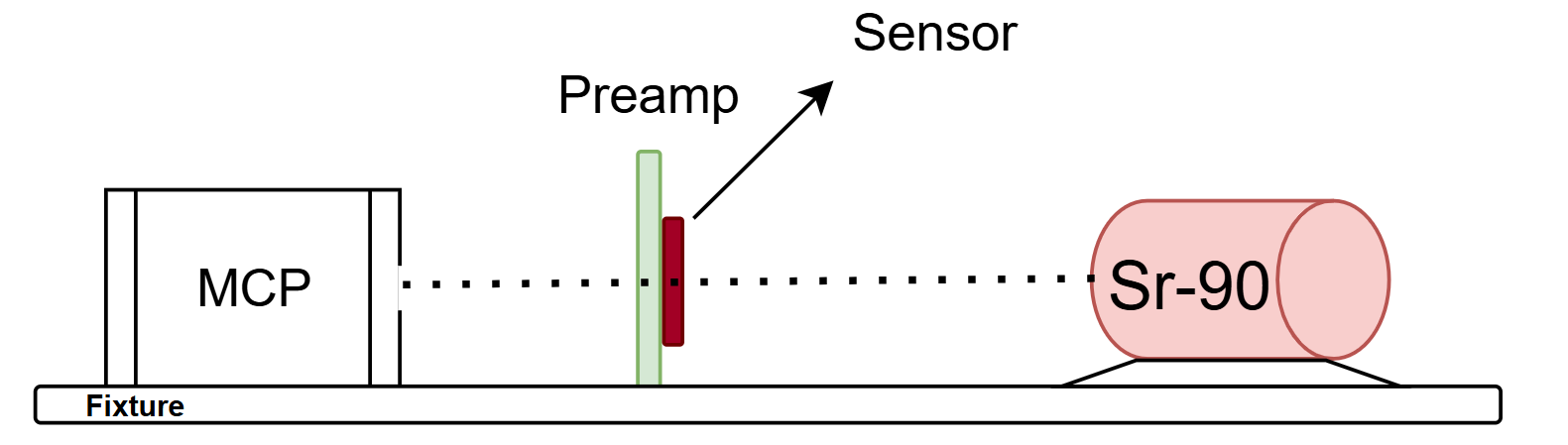}
	\caption{Setup of $\beta$ scope test.}
	\label{fig:setupofbeta-label}
\end{figure}

To evaluate the timing performance and SNR under practical operating conditions, a USTC-fabricated~\cite{ustclgad} sensor was bump-bonded to our electronics board and characterized using a \ce{^{90}Sr} beta radiation source . The sensor features an active thickness of \SI{50}{\micro\meter}, a pad area of \SI{1.34}{\milli\meter} $\times$ \SI{1.34}{\milli\meter}, and a measured capacitance of approximately \SI{4}{\pico\farad} at the operating bias voltage of \SI{-165}{\volt}. As illustrated in \autoref{fig:setupofbeta-label}, the entire apparatus was housed within a temperature-controlled environmental chamber equipped with electromagnetic shielding. The single-channel readout board under test, coupled with the device-under-test (DUT) detector, was positioned directly opposite the radiation source. A microchannel plate photomultiplier tube (MCP--PMT) served as the reference detector, exhibiting a time resolution of approximately 10 ps at \SI{-3000}{\volt}~\cite{bortfeldt2020163592}. The assembly was precisely center-aligned to ensure $\beta$-particles traversed the primary detector before reaching the reference detector. For comparative performance evaluation, an alternative preamplifier board was tested under identical conditions. In this study, the aforementioned reference amplifier board is designated as the USTC-V1 board~\cite{GE:Amp}.

All waveform data were acquired using the HDO9404 oscilloscope, algorithms processed the dual-channel waveforms to extract three critical parameters: collected charge, particle time-of-arrival (TOA), and TOA difference (${\Delta}$TOA). Timing determination employed a 50\% constant fraction discrimination (CFD) method to effectively mitigate time-walk effects. Given the substantial signal amplitude from the LGAD, a 100 mV trigger threshold was implemented, whereas the lower-amplitude MCP-PMT signal utilized a 10 mV threshold.Prior to measurements, the apparatus underwent thermal stabilization within the environmental chamber at 20 °C. The LGAD detector was biased at \qty{-165}{\volt}, while the MCP-PMT operated at -3000 V.

\begin{figure}[ht]
	\centering
	\includegraphics[width=0.5\linewidth]{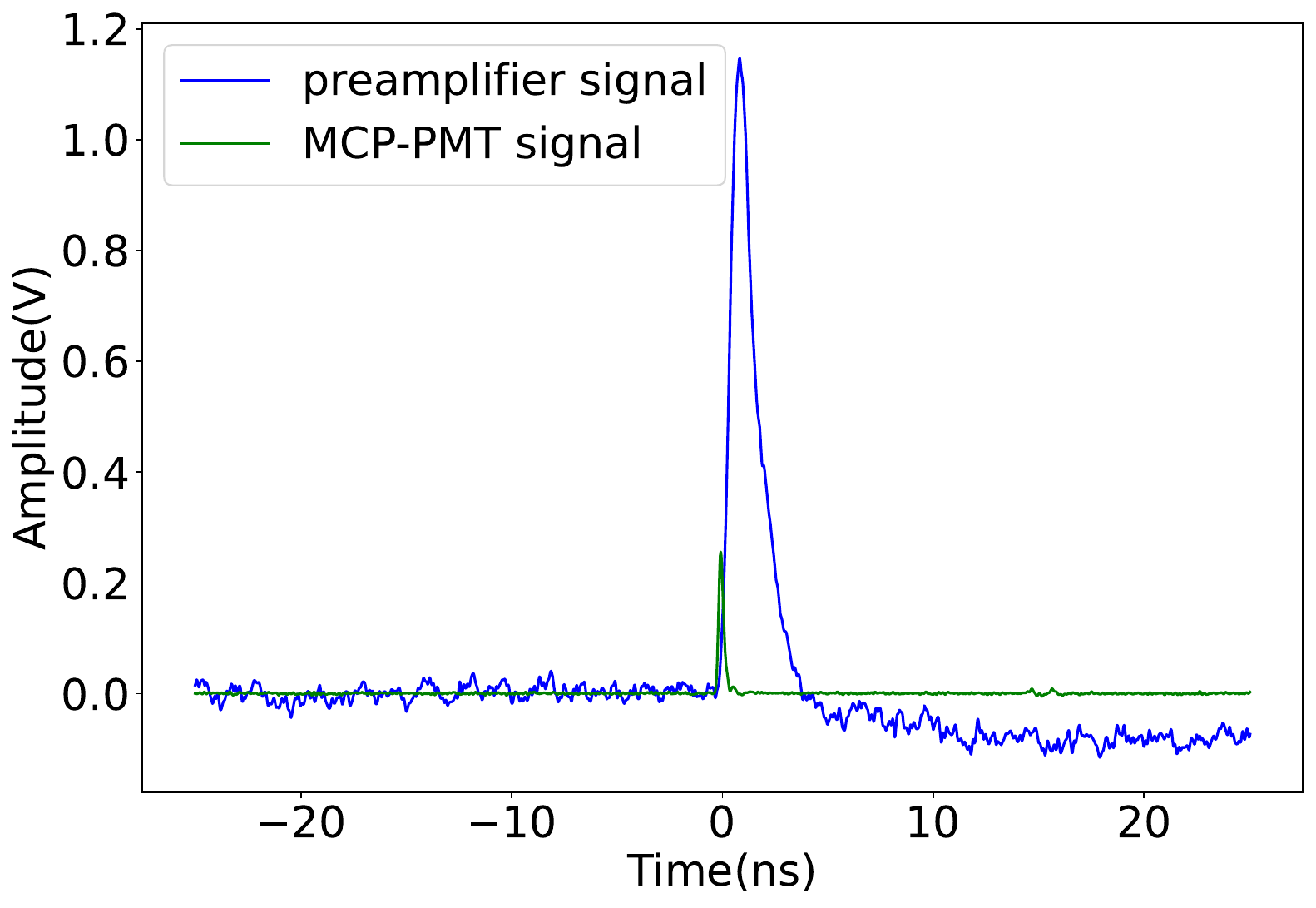}
	\caption{One representative output waveform from new preamplifier and MCP-PMT.}
	\label{fig:Representative_waveform-label}
\end{figure}

Representative waveforms obtained under these conditions are presented in \autoref{fig:Representative_waveform-label}. The observed undershoot in the signal from the amplification board is attributed to the use of AC coupling throughout the amplification circuit. The coupling capacitors accumulate charge during signal transmission, and this charge is subsequently released slowly through the associated resistors, leading to a gradual recovery of the signal back to its baseline level. This phenomenon is commonly referred to as baseline shift. Such drift may demand active baseline correction for high-rate operations. However, it poses no significant issue for the low-count-rate measurements reported here.

\begin{figure}[h]
    \centering

    \begin{minipage}{\textwidth}
        \centering
        \begin{subfigure}{0.25\textwidth}
            \includegraphics[width=\textwidth]{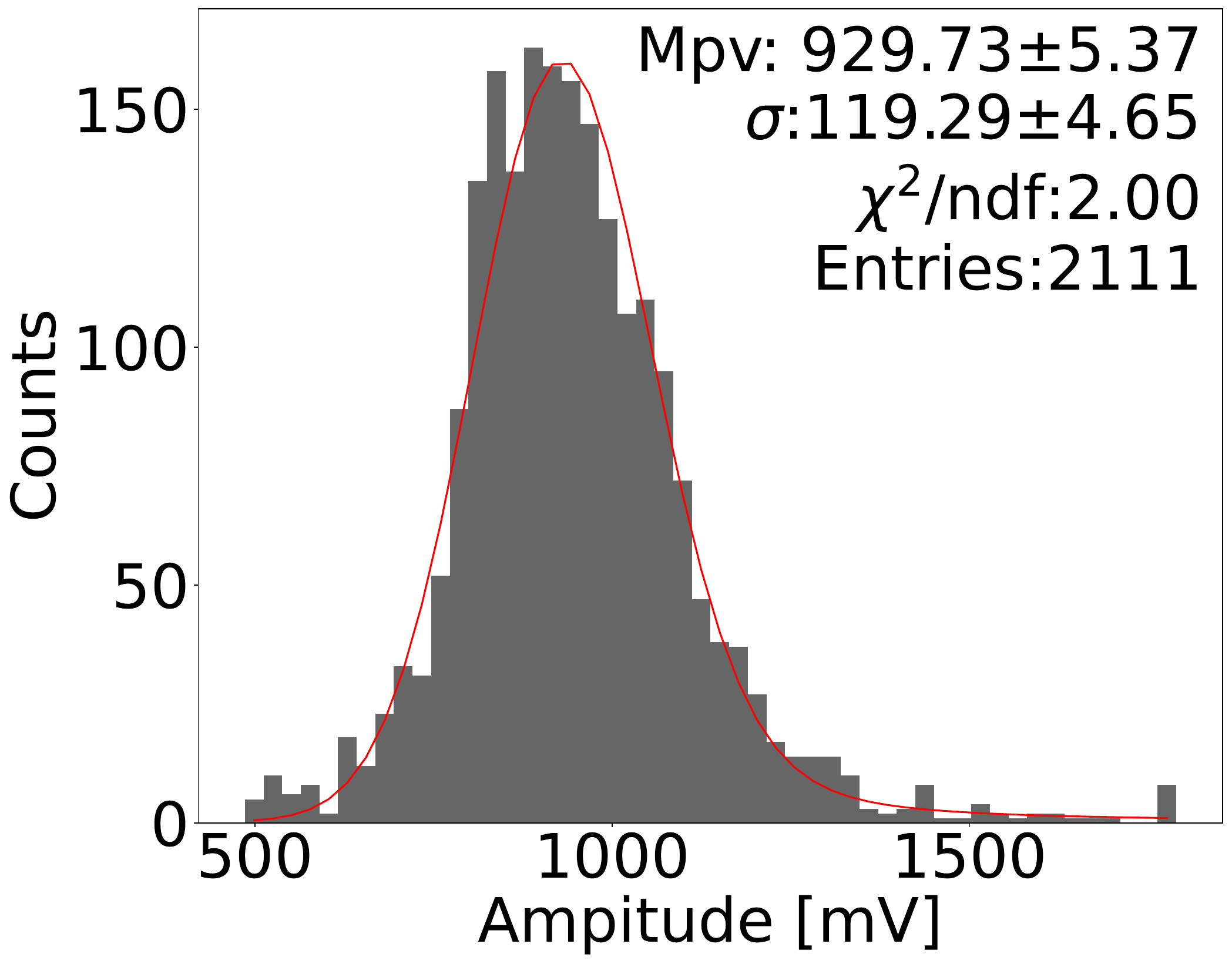}
            \caption{}
        \end{subfigure}\hspace{2em}%
        \begin{subfigure}{0.25\textwidth}
            \includegraphics[width=\textwidth]{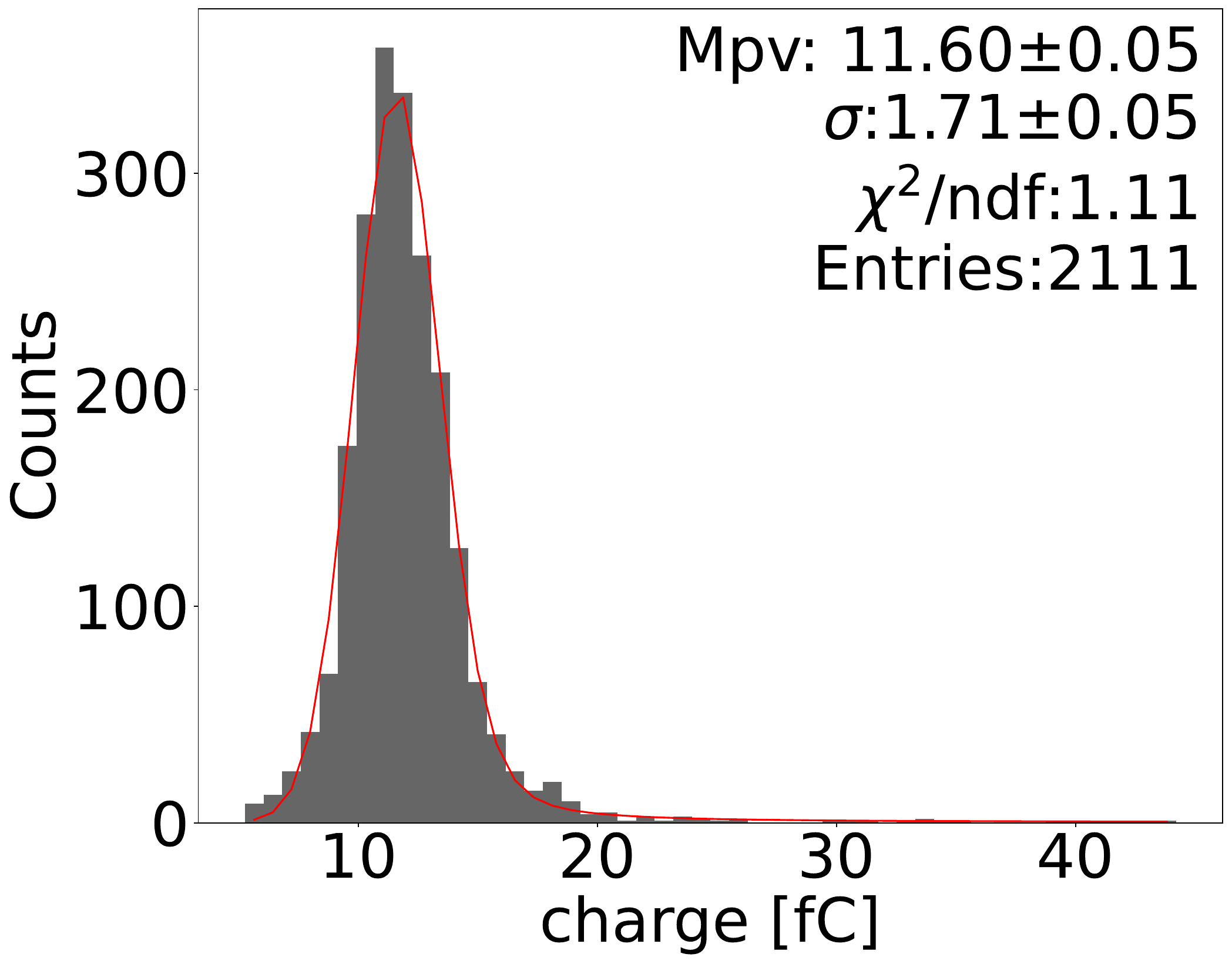}
            \caption{}
        \end{subfigure}\hspace{2em}%
        \begin{subfigure}{0.25\textwidth}
            \includegraphics[width=\textwidth]{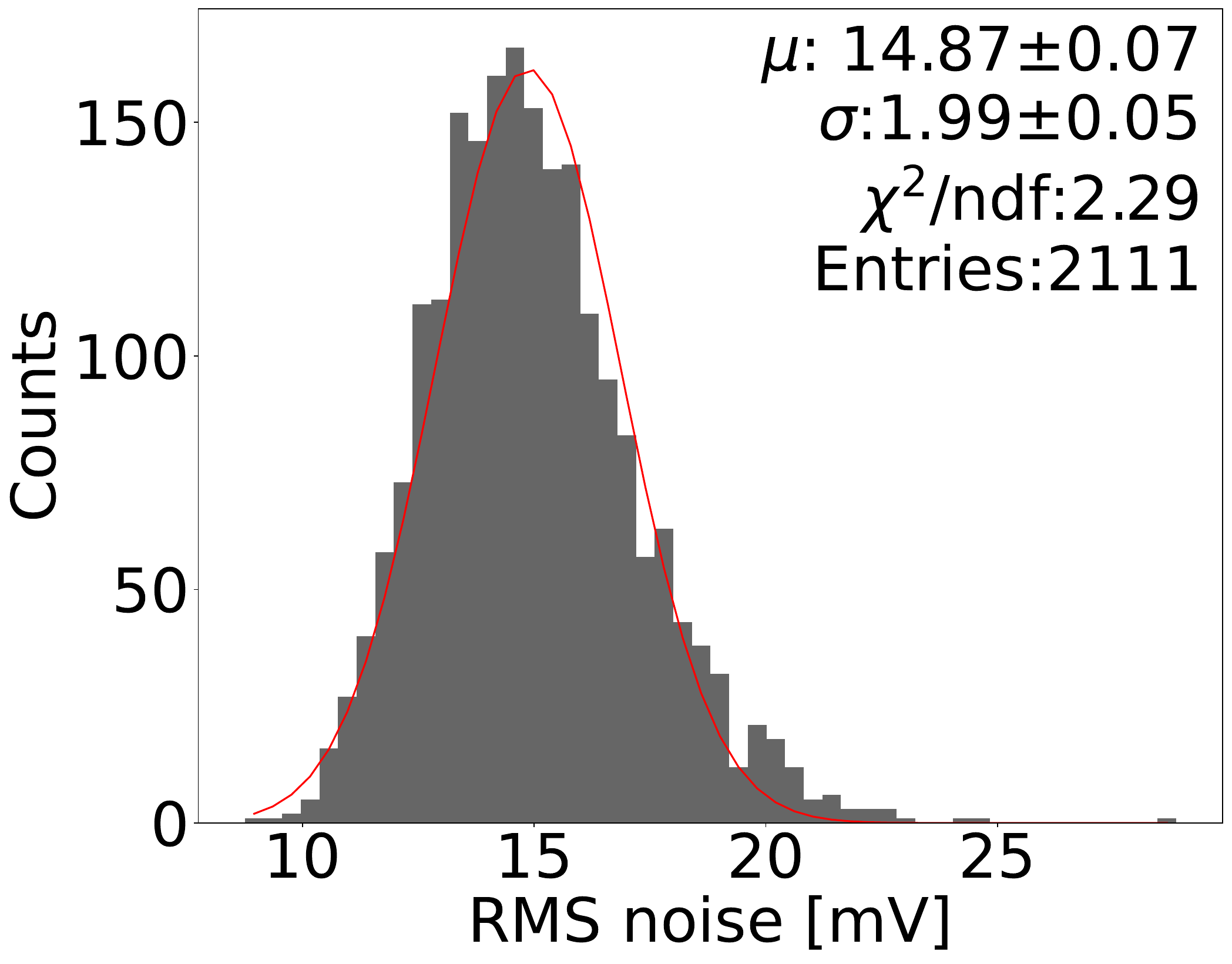}
            \caption{}
        \end{subfigure}
    \end{minipage}

    \vspace{0.5em}

    \begin{minipage}{\textwidth}
        \centering
        \begin{subfigure}{0.25\textwidth}
            \includegraphics[width=\textwidth]{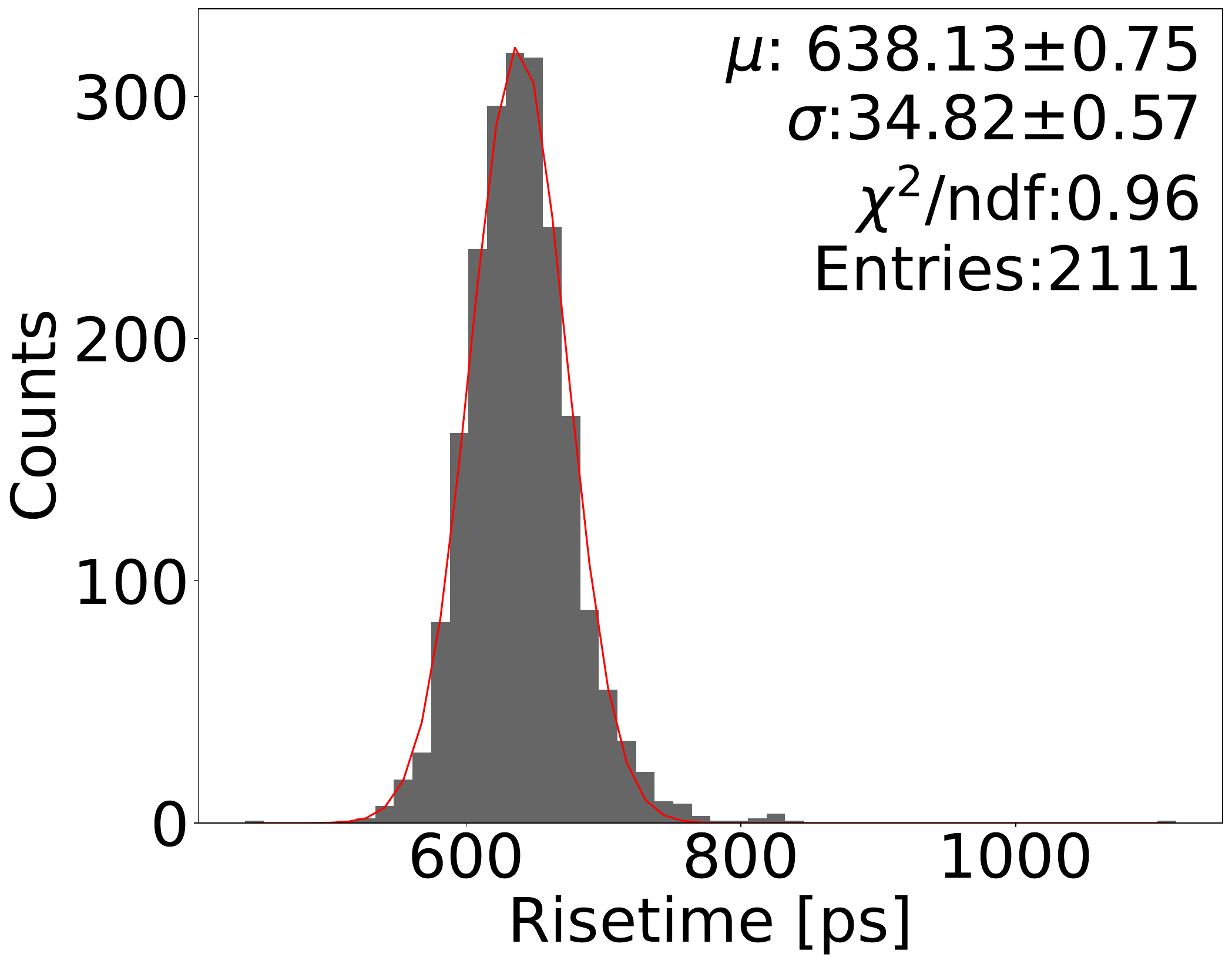}
            \caption{}
        \end{subfigure}\hspace{2em}%
        \begin{subfigure}{0.25\textwidth}
            \includegraphics[width=\textwidth]{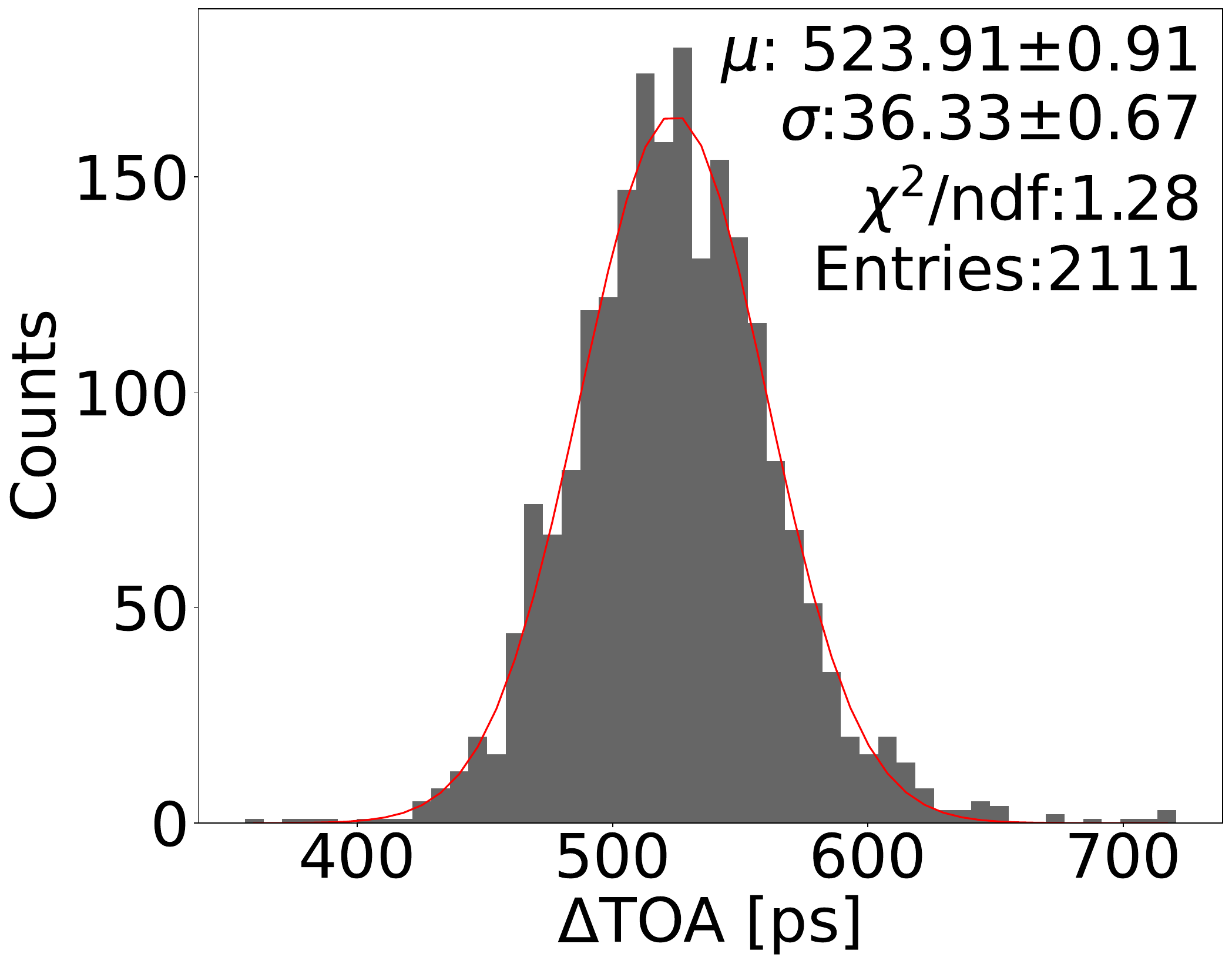}
            \caption{}
        \end{subfigure}
    \end{minipage}

    \vspace{0.5em}

    \begin{minipage}{\textwidth}
        \centering
        \begin{subfigure}{0.25\textwidth}
            \includegraphics[width=\textwidth]{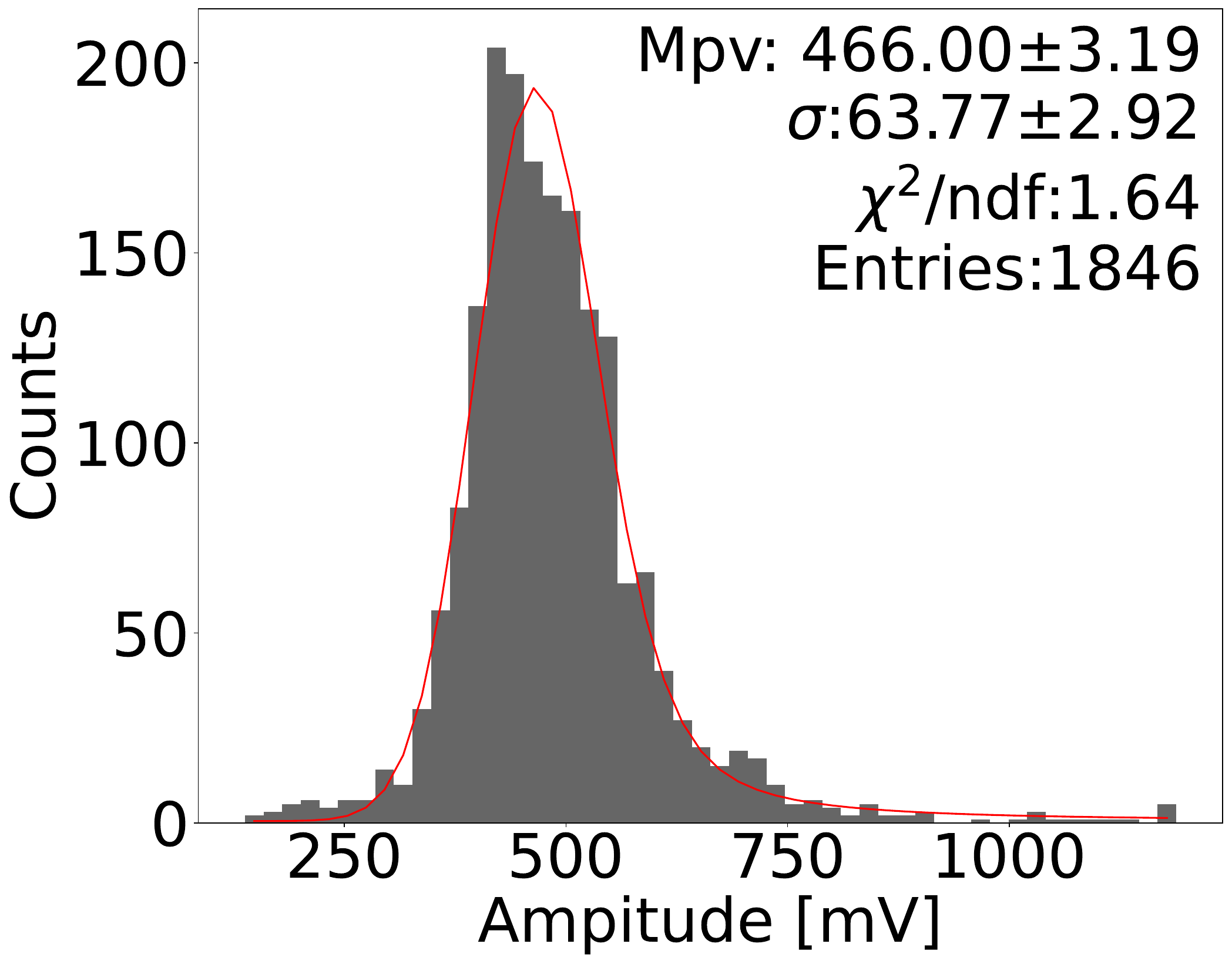}
            \caption{}
        \end{subfigure}\hspace{2em}%
        \begin{subfigure}{0.25\textwidth}
            \includegraphics[width=\textwidth]{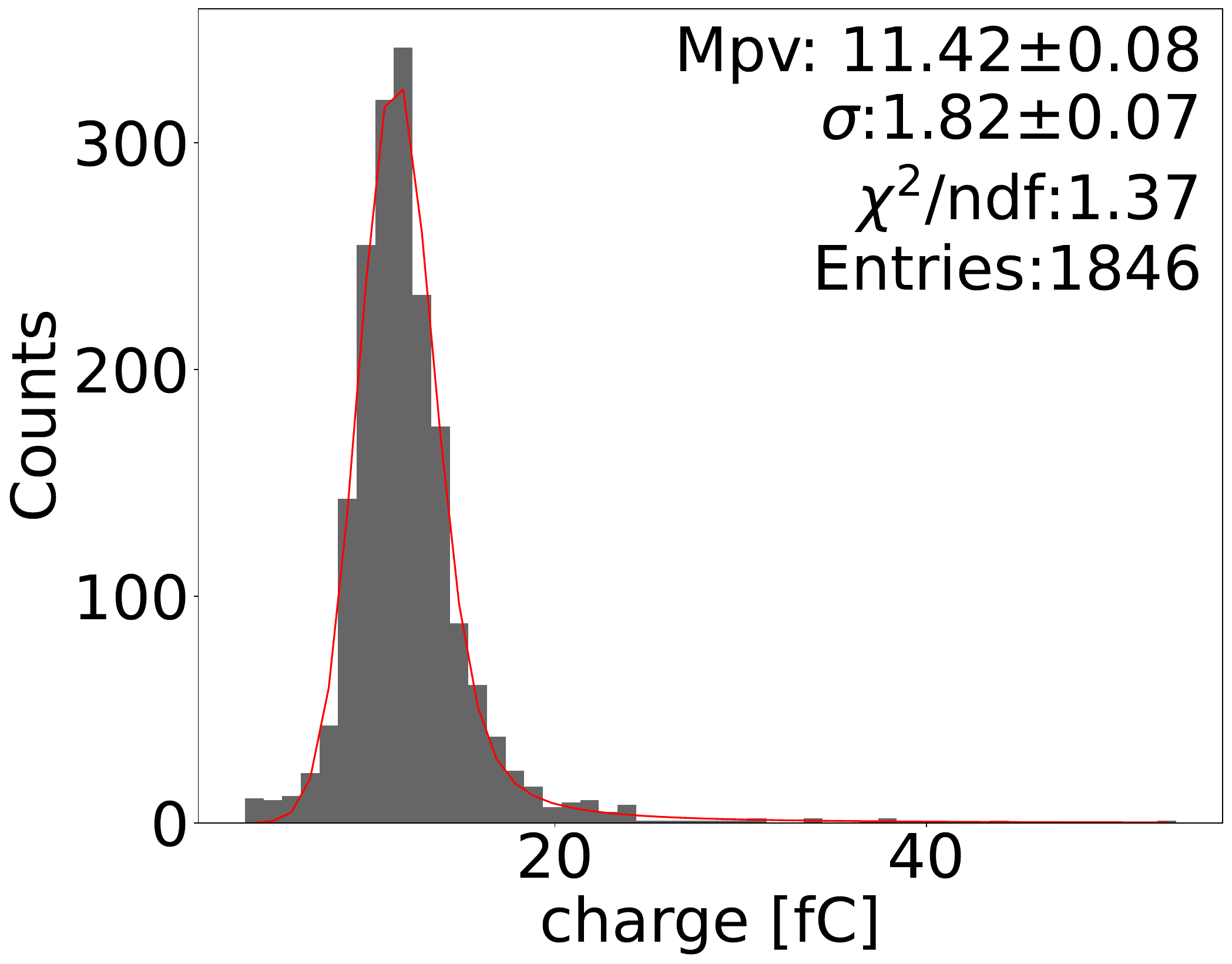}
            \caption{}
        \end{subfigure}\hspace{2em}%
        \begin{subfigure}{0.25\textwidth}
            \includegraphics[width=\textwidth]{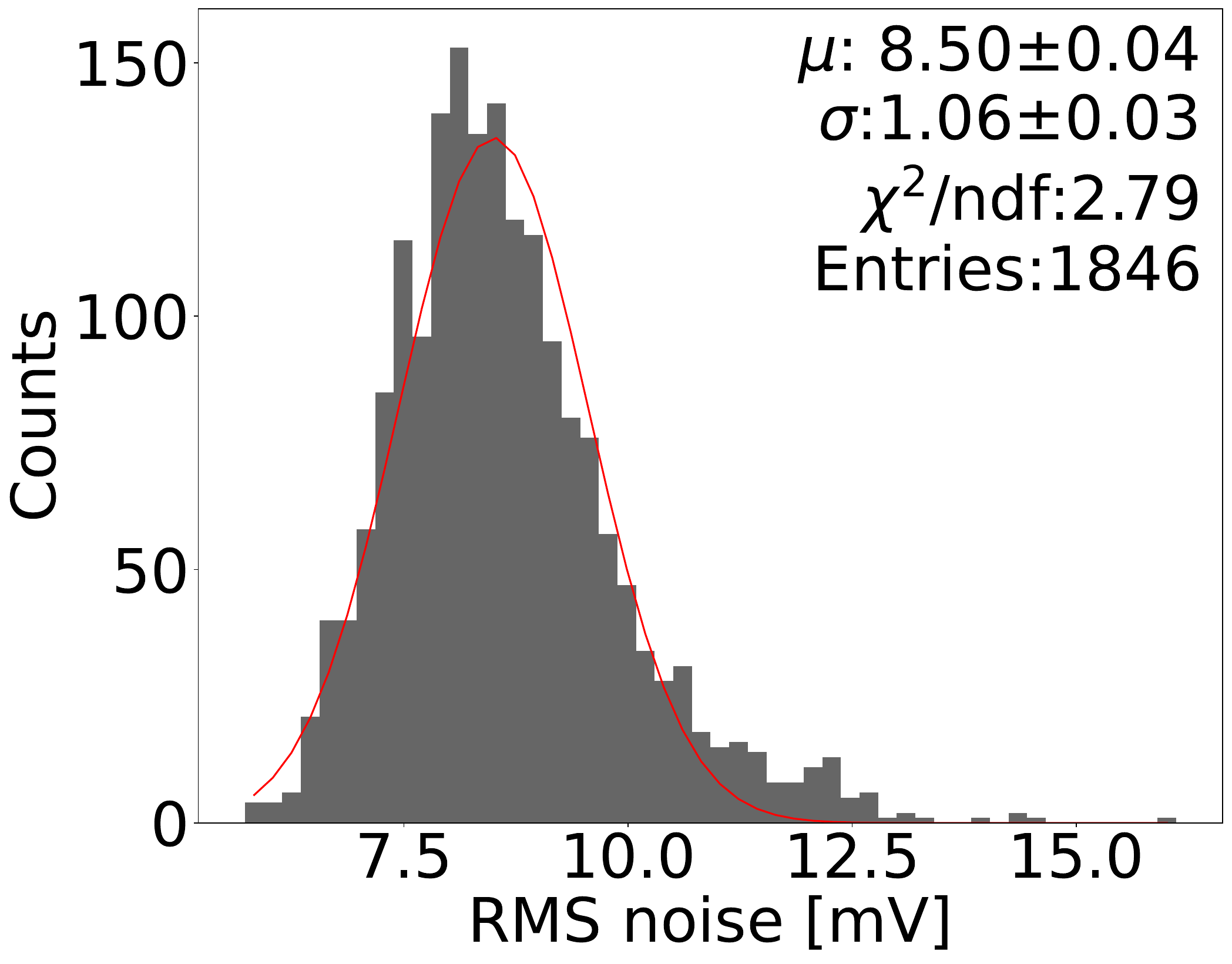}
            \caption{}
        \end{subfigure}
    \end{minipage}

    \vspace{0.5em}

    \begin{minipage}{\textwidth}
        \centering
        \begin{subfigure}{0.25\textwidth}
            \includegraphics[width=\textwidth]{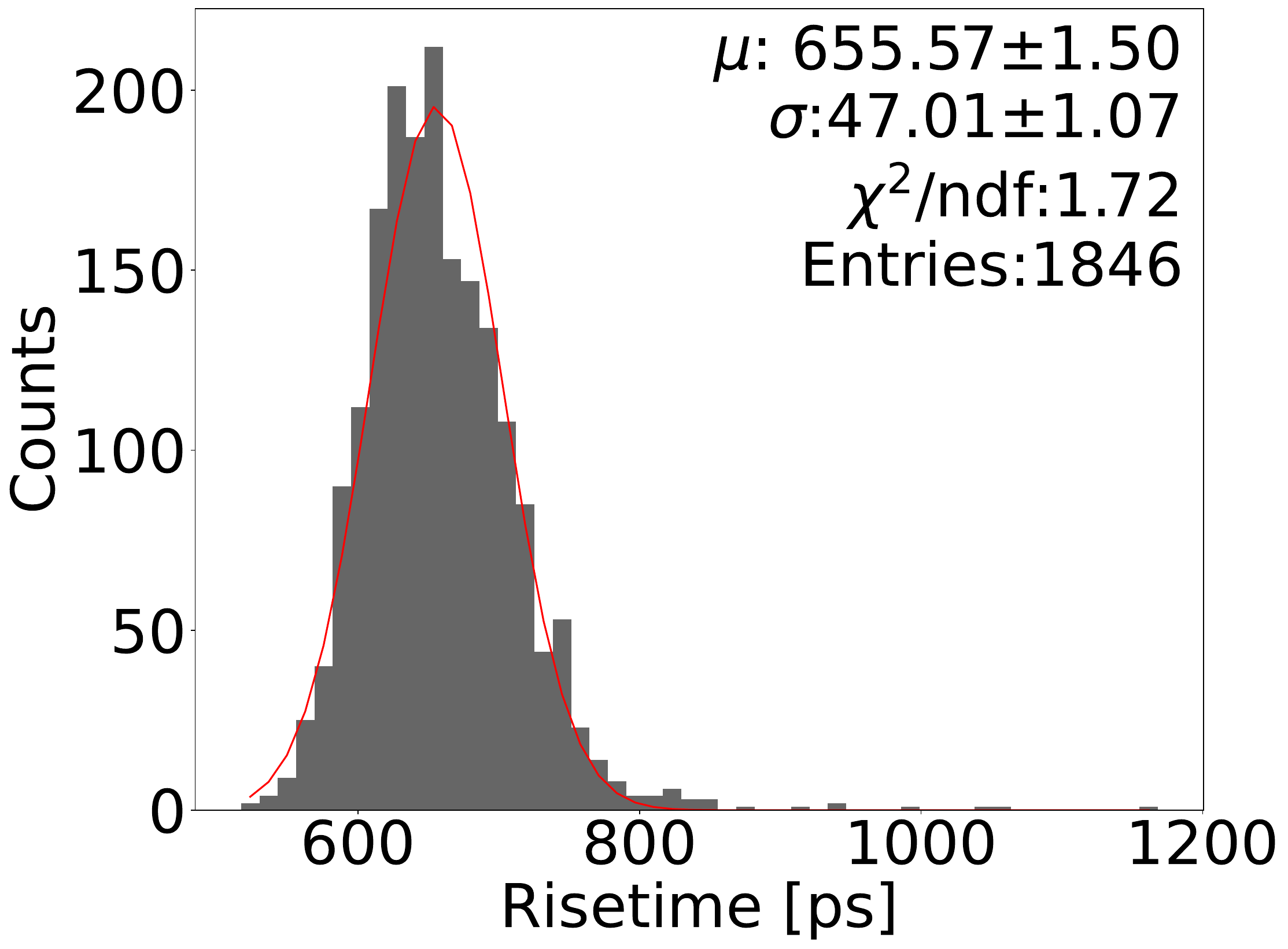}
            \caption{}
        \end{subfigure}\hspace{2em}%
        \begin{subfigure}{0.25\textwidth}
            \includegraphics[width=\textwidth]{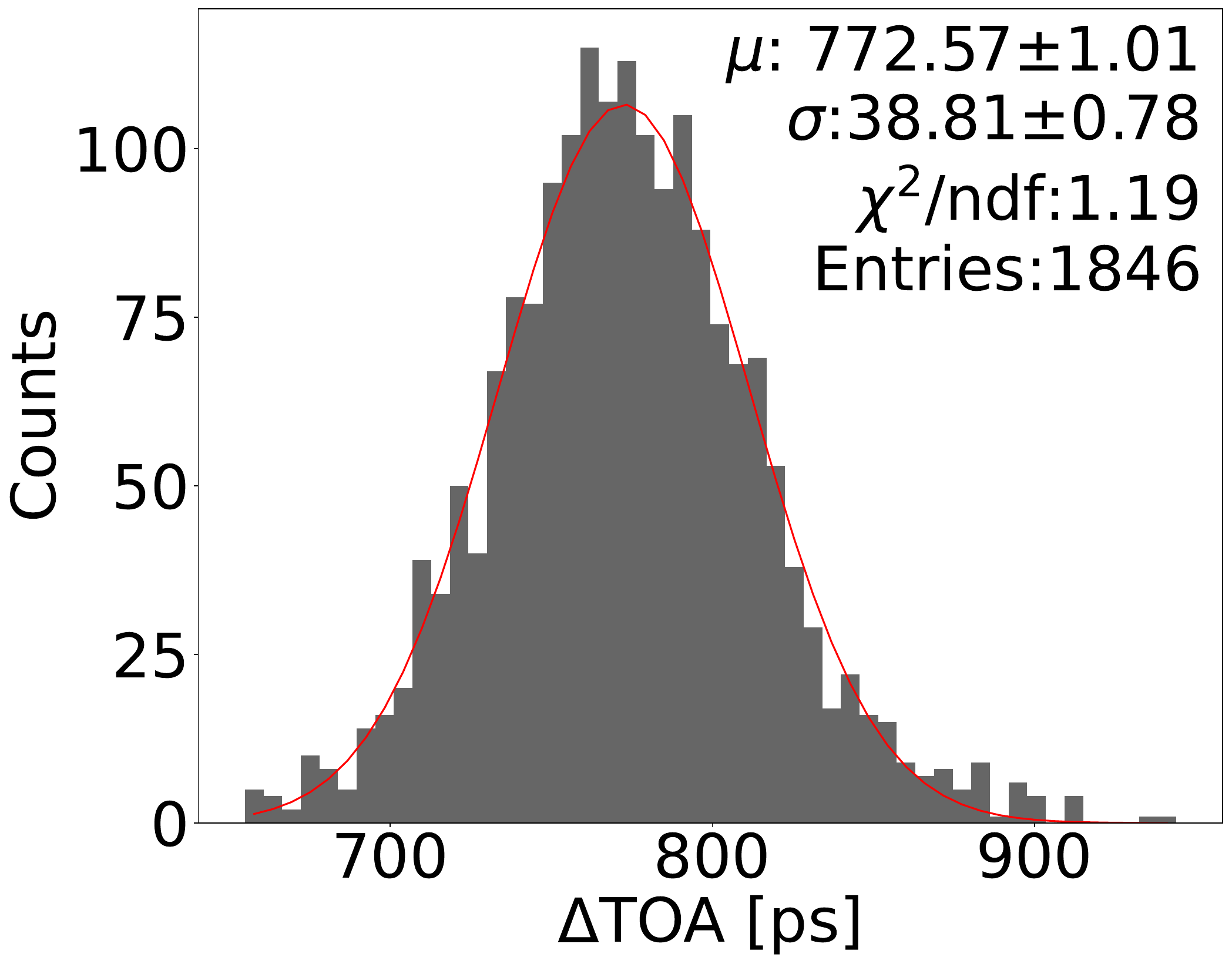}
            \caption{}
        \end{subfigure}
    \end{minipage}

	\caption{Distributions of output amplitude (a), collected charges (b), RMS noise (c), risetime (d) and ${\Delta}$TOA (e) measured with our single-channel boards and corresponding results (f--j) are measured with USTC-V1 boards.}
	\label{fig:two_rows}
\end{figure}

Following extended data acquisition, statistical analysis yielded the target parameters: amplitude, collected charge, ${\Delta}$TOA distribution, and detector timing resolution. Additionally, RMS noise characteristics were quantified from baseline segments of the pre-trigger waveform samples~\cite{li2022167008}. The comprehensive test results are summarized in \autoref{fig:two_rows}.

The most probable value (MPV) of the signal amplitude ($V_{\text{om}}$) was measured as 929.73 mV, while the root mean square (RMS) value of the board noise ($V_{\text{n}}$) was determined to be 14.87 mV. Using
\begin{equation}
	\mathrm{SNR} =\frac{V_{\text{om}}}{V{\text{n}}},
    \label{eq:SNR}
\end{equation}
the corresponding  SNR was calculated as 62.52. While the MPV of collected charge (Q) is 11.60 fC, yielding an ENC of 0.19 fC based on:
\begin{equation}
	\mathrm{ENC} = \frac{Q}{\mathrm{SNR}}.
\end{equation}
Under identical test conditions, analysis of the reference amplification board produced an SNR of 54.82 and an ENC of 0.21 fC.

It is evident that the readout board under evaluation exhibits a higher gain compared with the reference preamplifier board. Although this enhancement is accompanied by an increase in RMS noise, the overall SNR is improved. This improvement is directly reflected in the timing resolution. The measured combined DUT--MCP-PMT resolutions of 38.81 ps and 36.33 ps include the contribution of the MCP-PMT reference detector ($\leq$ 10 ps intrinsic resolution). Deconvolving this contribution using the conservative upper bound of 10 ps yields intrinsic timing resolutions of 37.50 ps (USTC-V1 board) and 34.93 ps (our board), respectively. The deconvolved intrinsic values are reported as the final timing resolutions, showing an improvement from 37.50 ps to 34.93 ps under identical experimental conditions.

To further characterize the readout performance, the electronic jitter is extracted using
\begin{equation}
	\sigma _{\mathrm{jitter}} = \frac{V_{\text{n}}}{(V_{\text{om}}/t_r)}=\frac{t_r}{\mathrm{SNR}}  \label{eq:jitter}
\end{equation}
to extract the electronic jitter~\cite{WOS:P06008}. The results are 10.9 ps for the preamplifier board and 12.6 ps for the USTC‑V1 board.

The relatively modest difference in timing resolution can be attributed to the inherently high internal gain of the LGAD sensor itself. For a minimum-ionizing particle depositing approximately 0.5\,fC in the 50\,$\mu$m active layer, the measured collected charge of 11.60\,fC corresponds to an internal gain of about $11.60 / 0.5 \approx 23.2$. At this gain level, the LGAD generates a sufficiently large output signal that the timing performance becomes less sensitive to moderate variations in front-end electronics gain and noise. Consequently, even with a measurable improvement in SNR due to the higher-gain readout architecture, the ultimate timing resolution remains constrained primarily by the detector's intrinsic characteristics and the statistical fluctuations in charge collection, rather than by the electronic noise floor alone. This observation aligns with established principles in fast timing detector systems, where once the electronics noise is sufficiently suppressed relative to the signal amplitude, further gains yield diminishing returns in timing precision.

\subsubsection{Test results with PIN}

To verify the compatibility of our amplifier board with low charge detectors, we employed a PIN diode fabricated using the same process as the USTC-IME LGAD wafer but without internal gain layer as DUT. The PIN diode shares the same active thickness of \SI{50}{\micro\meter} and a comparable capacitance of approximately \SI{4}{\pico\farad} as the LGAD at the operating bias voltage. For reference, an LGAD detector coupled to the USTC-V1 amplifier board was used as the trigger channel. Since the PIN diode lacks internal gain, its signal amplitude when bonded directly to the USTC-V1 board proved insufficient for adequate  SNR. To address this, the feedback resistor of the preamplifier stage was adjusted to 1.13 k$\Omega$  to enhance the overall gain.



\begin{figure}[h]
    \centering

    \begin{minipage}{\textwidth}
        \centering
        \begin{subfigure}{0.25\textwidth}
            \includegraphics[width=\textwidth]{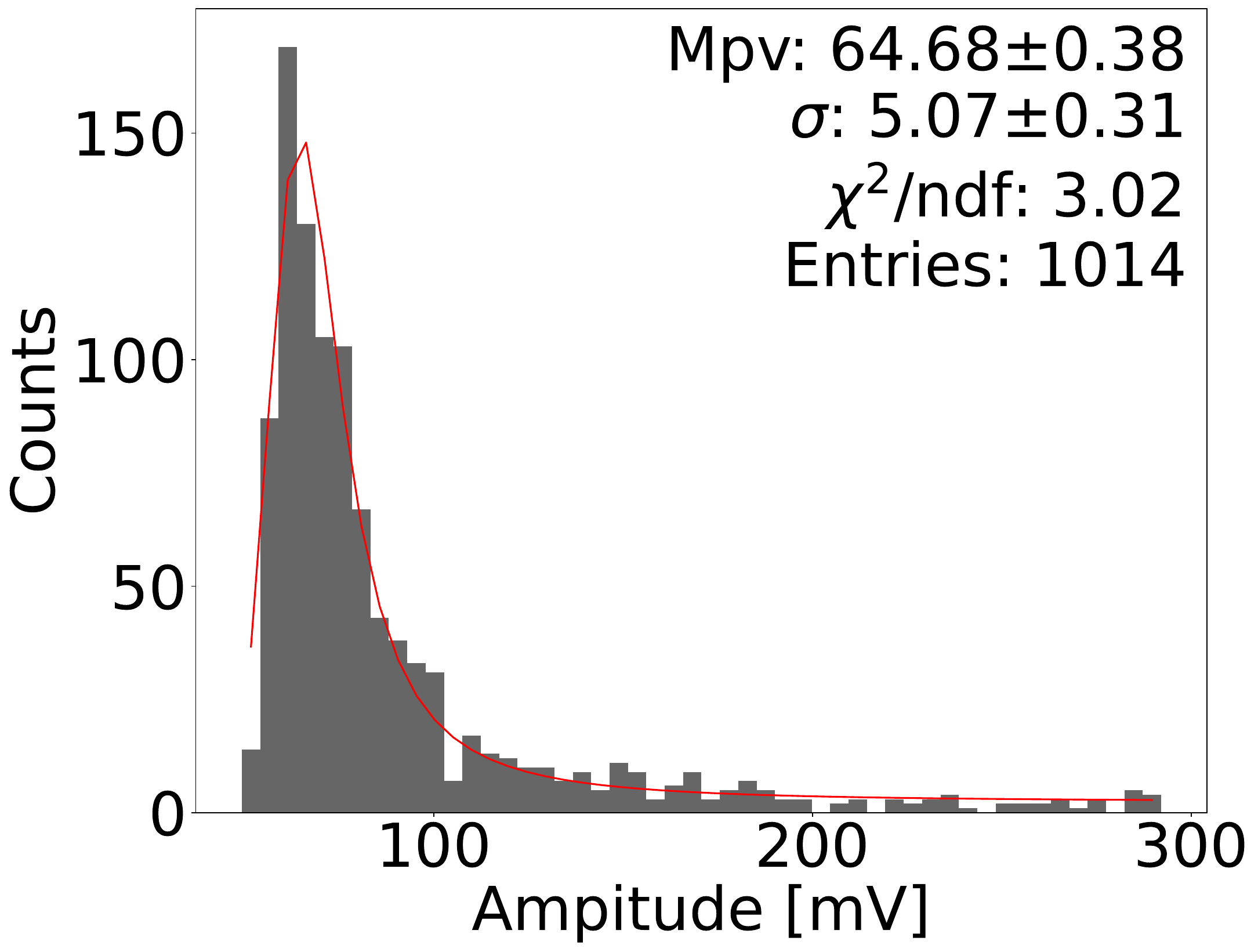}
            \caption{}
        \end{subfigure}\hspace{2em}%
        \begin{subfigure}{0.25\textwidth}
            \includegraphics[width=\textwidth]{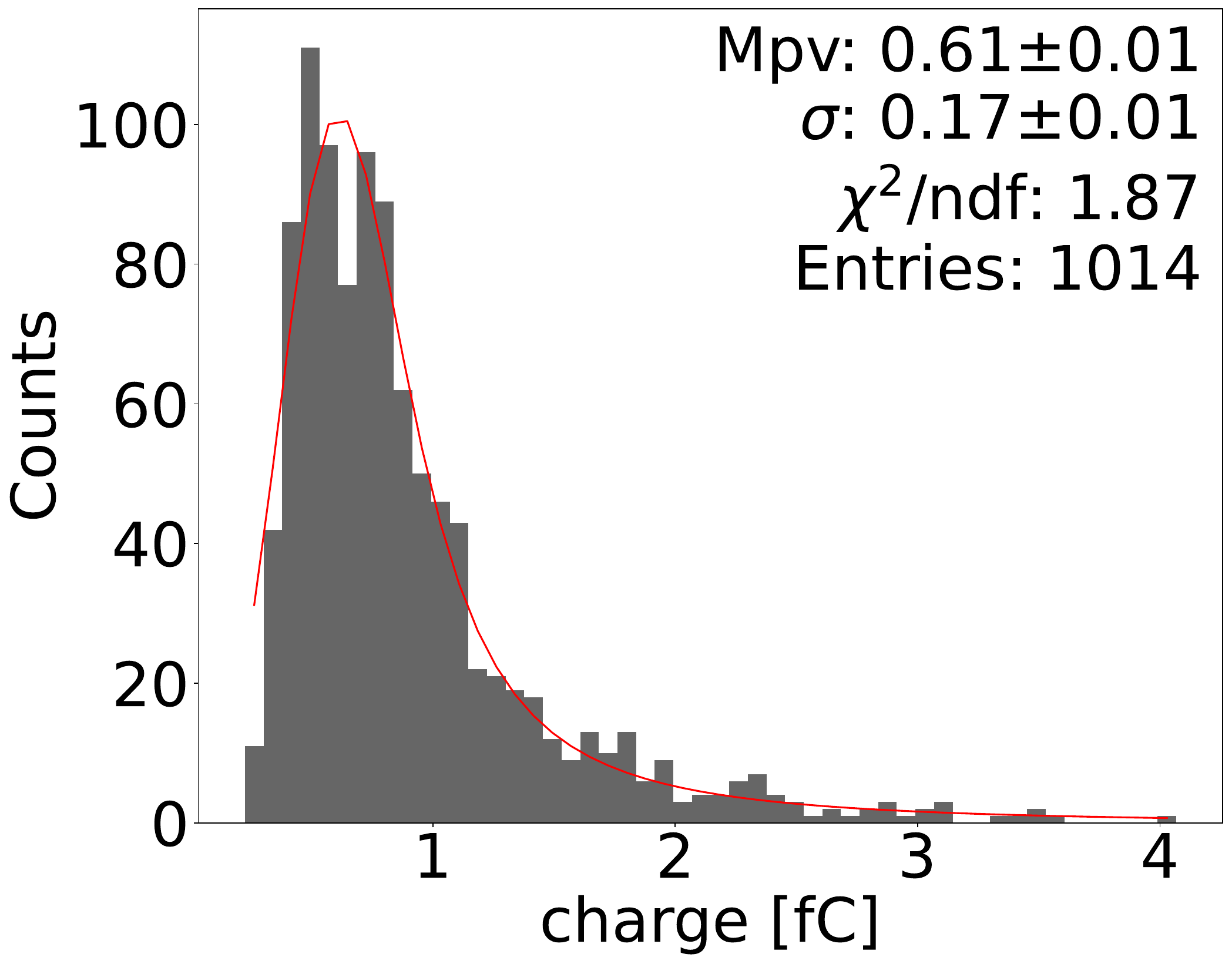}
            \caption{}
        \end{subfigure}\hspace{2em}%
        \begin{subfigure}{0.25\textwidth}
            \includegraphics[width=\textwidth]{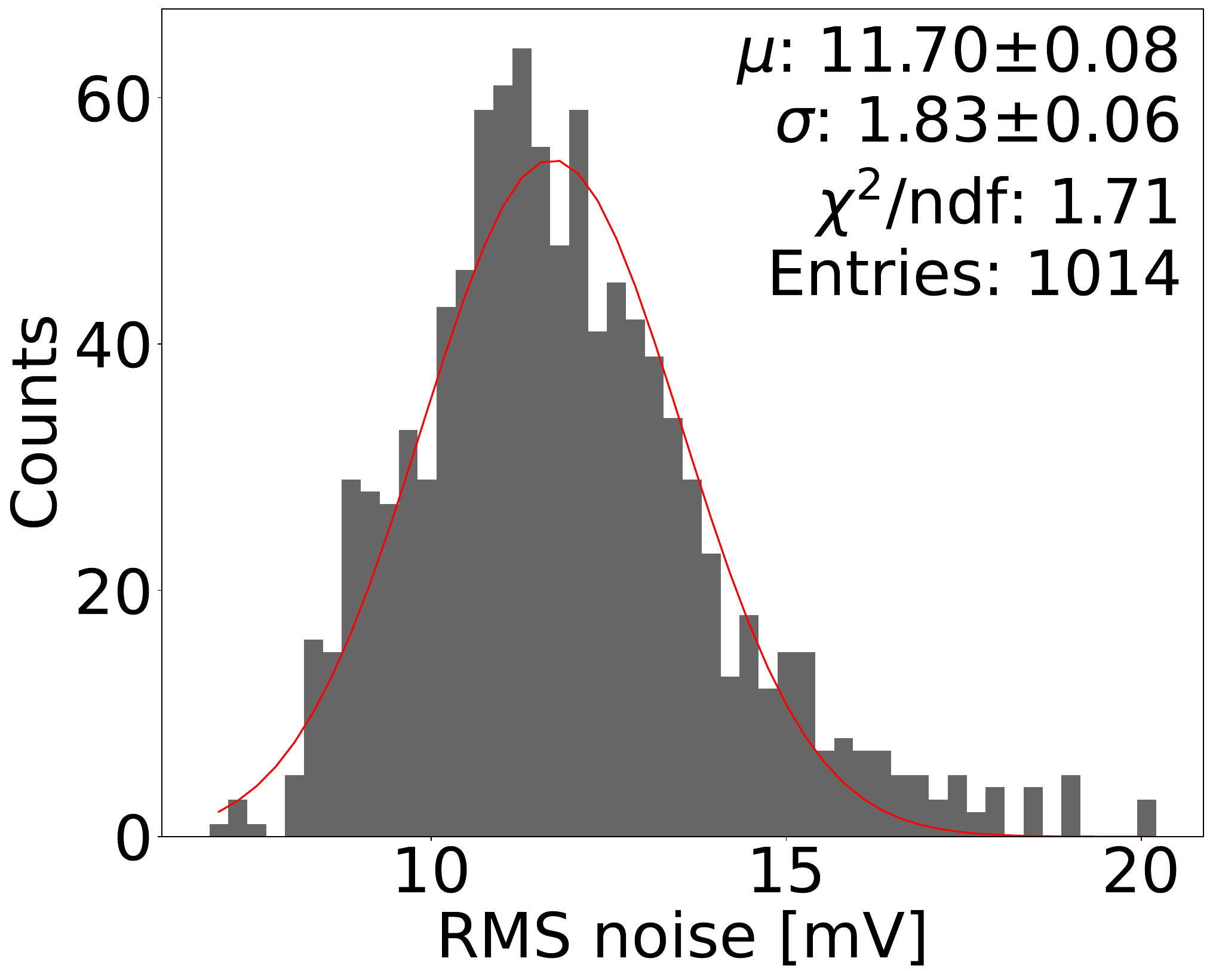}
            \caption{}
        \end{subfigure}
    \end{minipage}

    \vspace{0.5em}

    \begin{minipage}{\textwidth}
        \centering
        \begin{subfigure}{0.25\textwidth}
            \includegraphics[width=\textwidth]{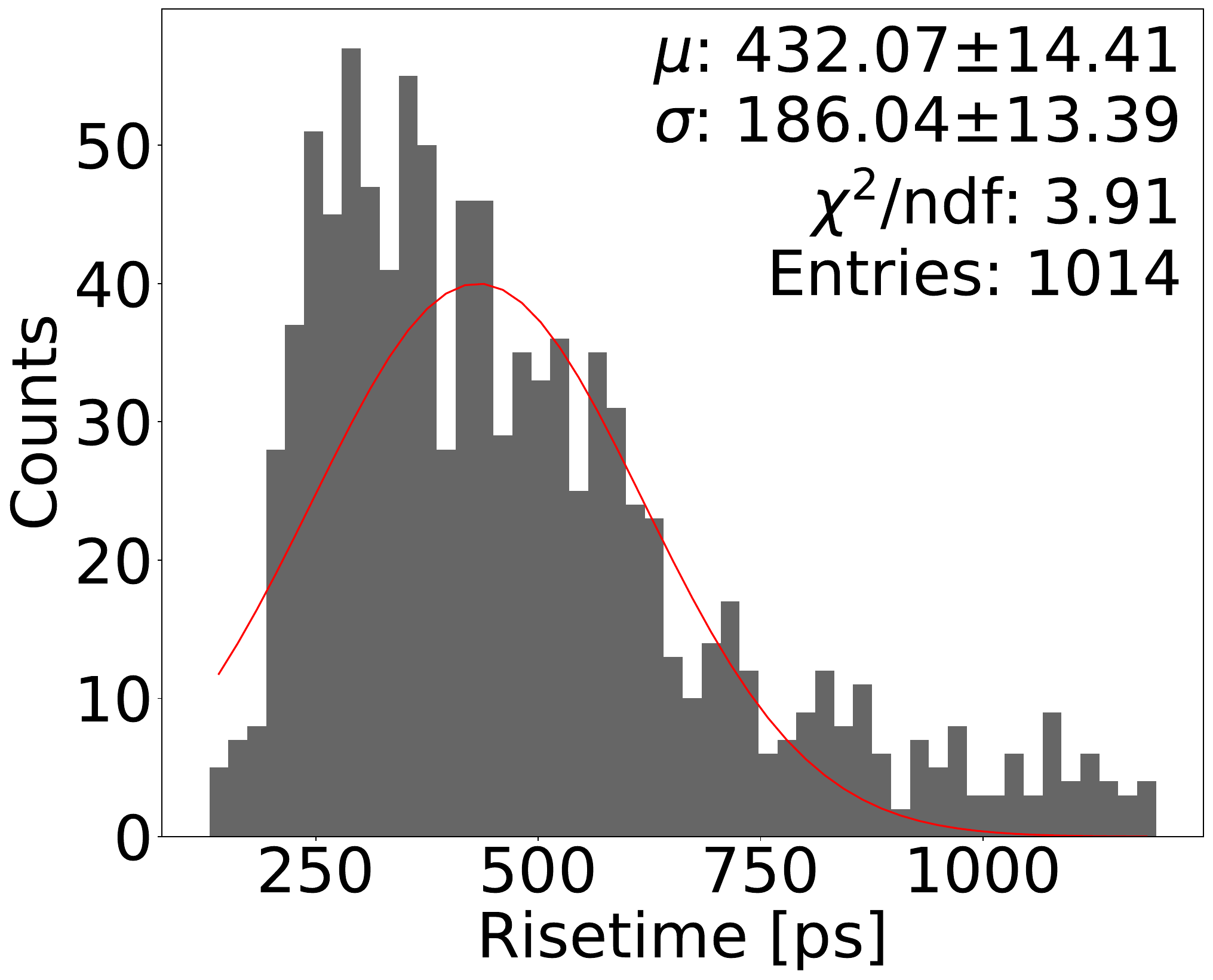}
            \caption{}
        \end{subfigure}\hspace{2em}%
        \begin{subfigure}{0.25\textwidth}
            \includegraphics[width=\textwidth]{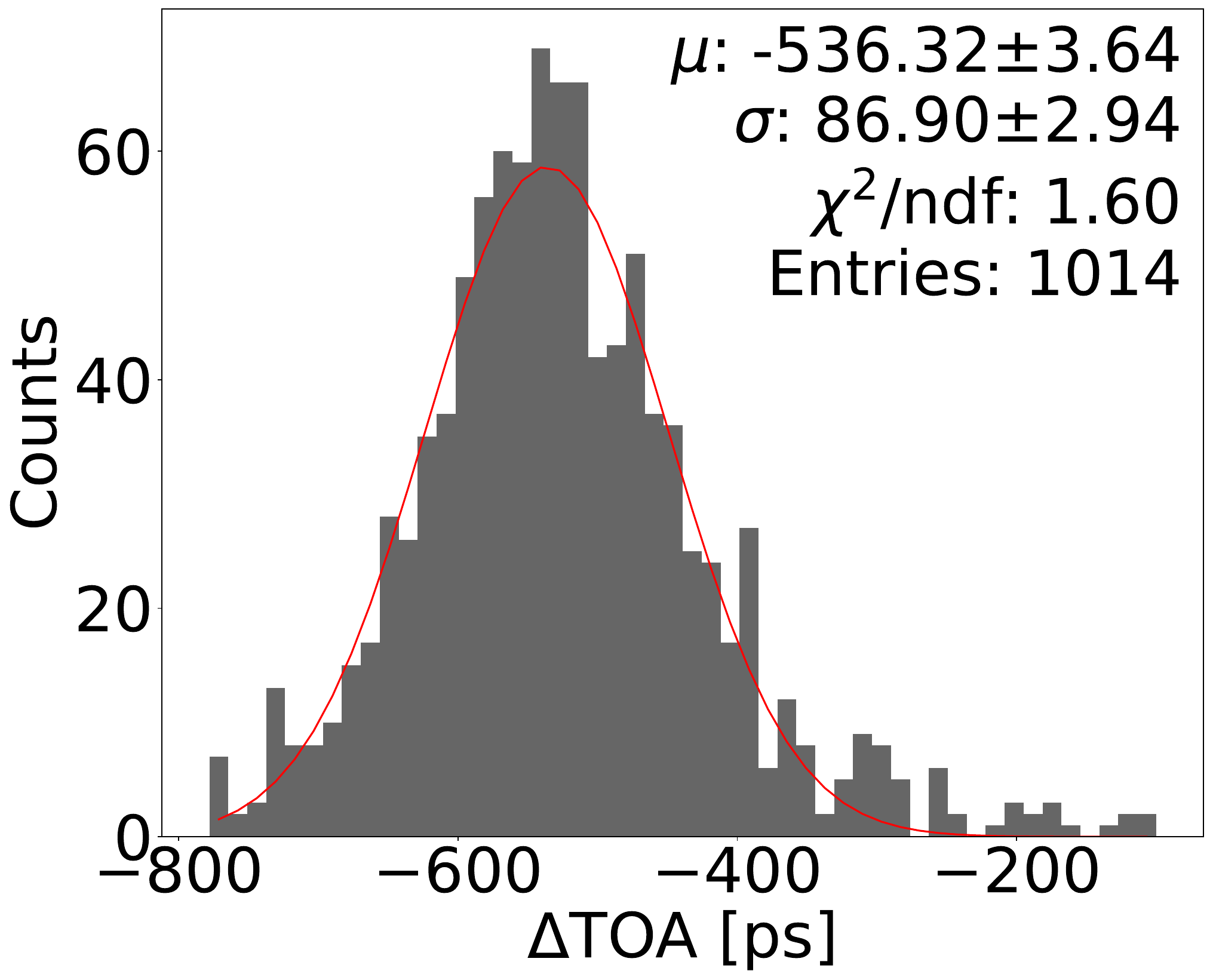}
            \caption{}
        \end{subfigure}
    \end{minipage}

    \vspace{0.5em}

    \begin{minipage}{\textwidth}
        \centering
        \begin{subfigure}{0.25\textwidth}
            \includegraphics[width=\textwidth]{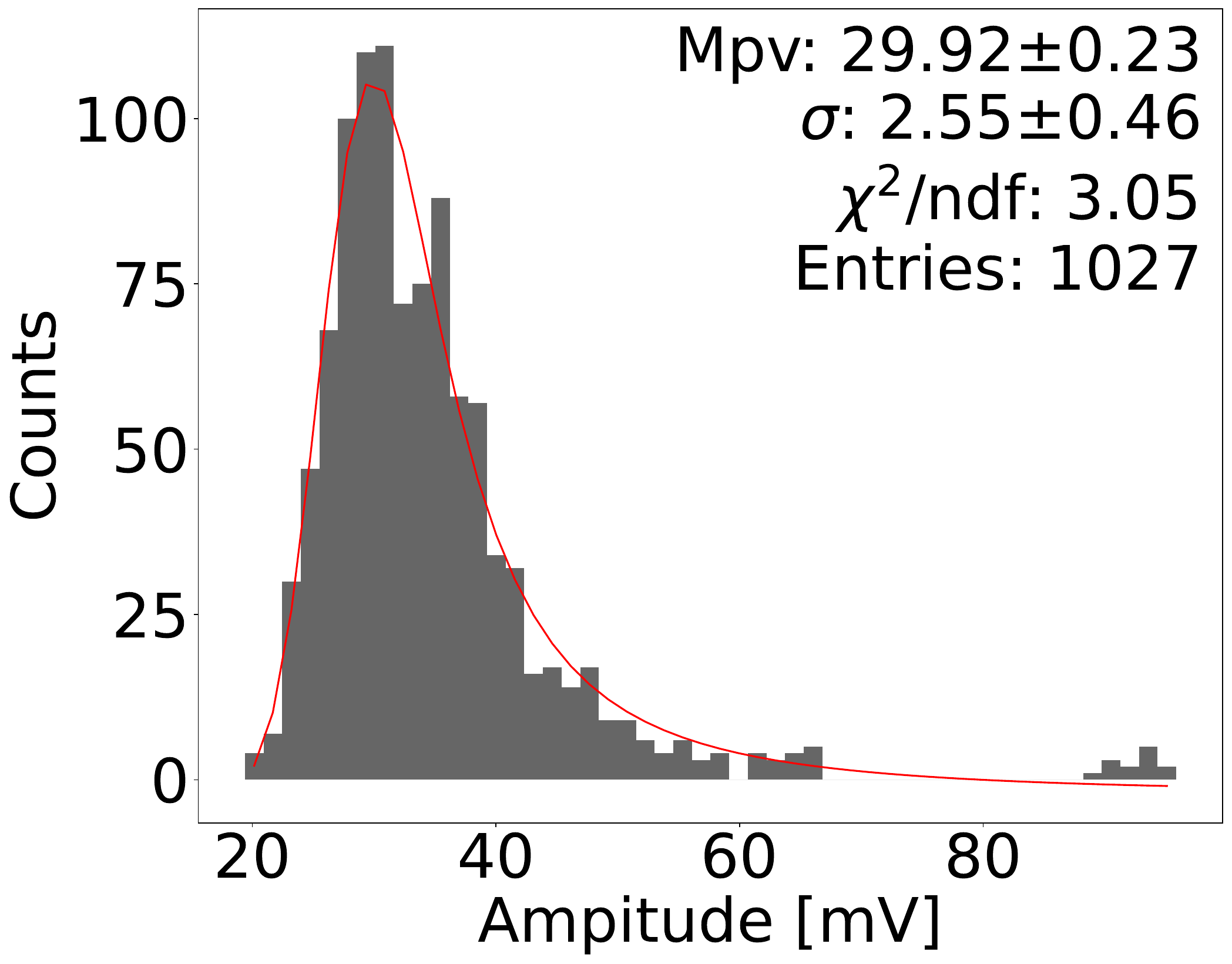}
            \caption{}
        \end{subfigure}\hspace{2em}%
        \begin{subfigure}{0.25\textwidth}
            \includegraphics[width=\textwidth]{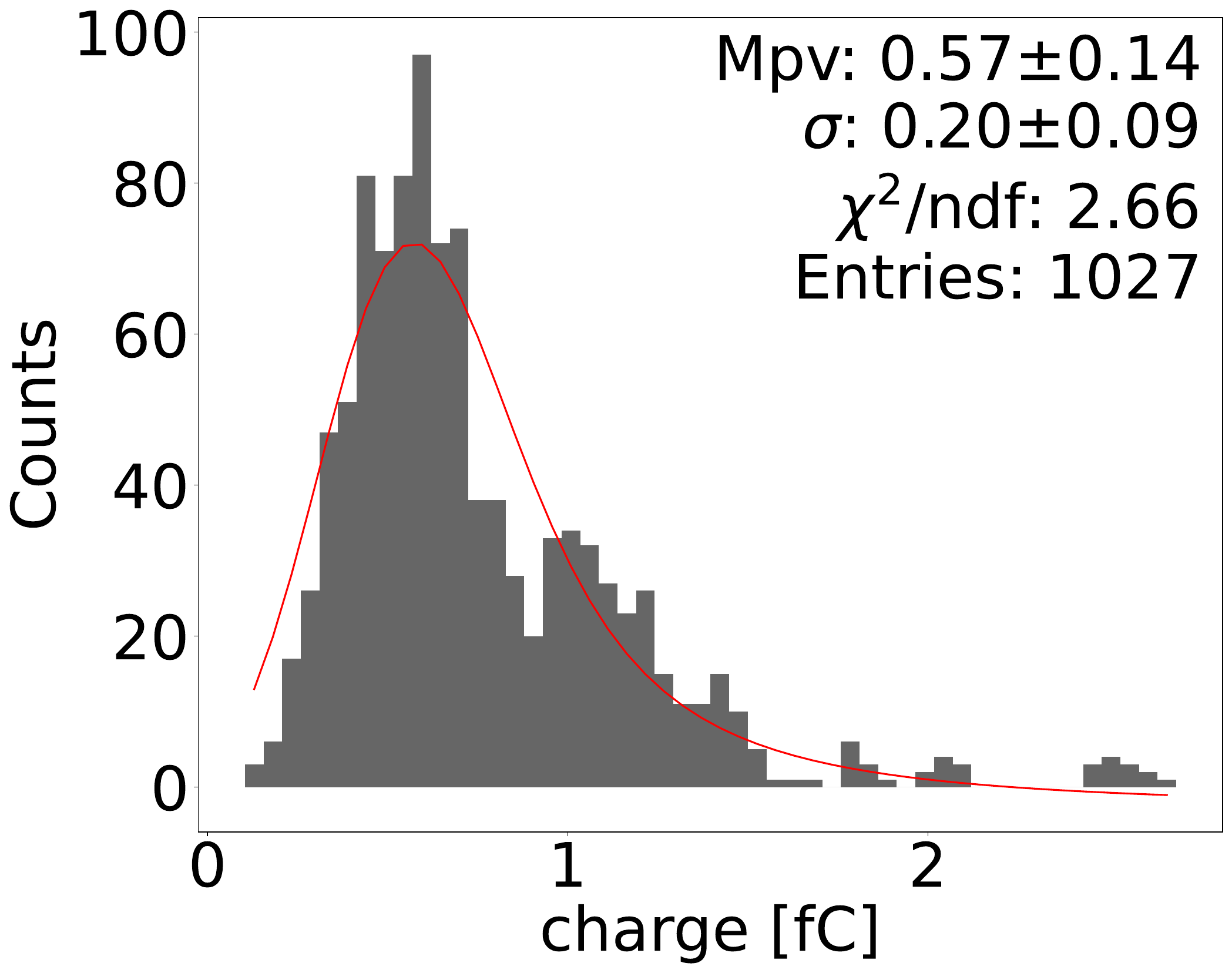}
            \caption{}
        \end{subfigure}\hspace{2em}%
        \begin{subfigure}{0.25\textwidth}
            \includegraphics[width=\textwidth]{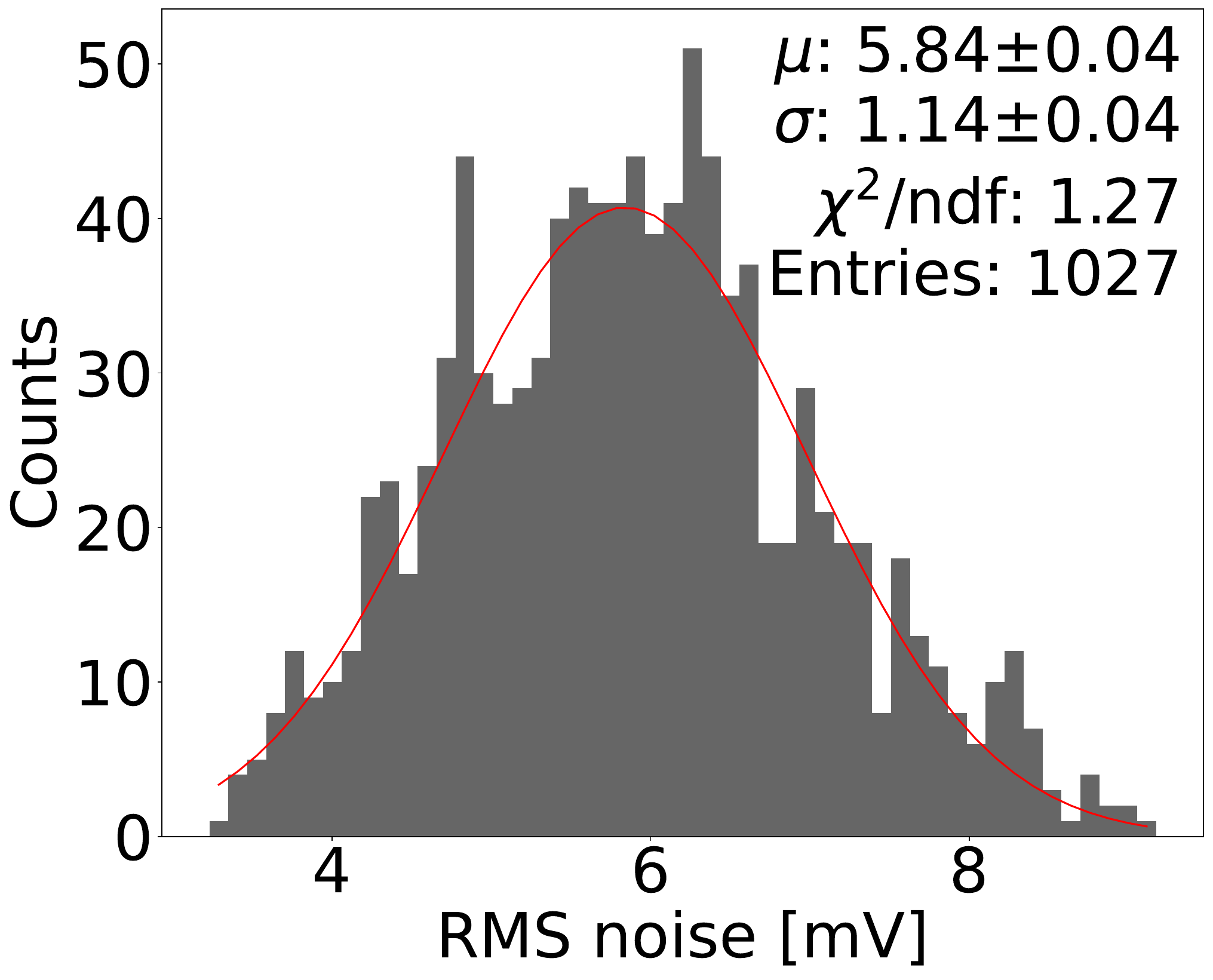}
            \caption{}
        \end{subfigure}
    \end{minipage}

    \vspace{0.5em}

    \begin{minipage}{\textwidth}
        \centering
        \begin{subfigure}{0.25\textwidth}
            \includegraphics[width=\textwidth]{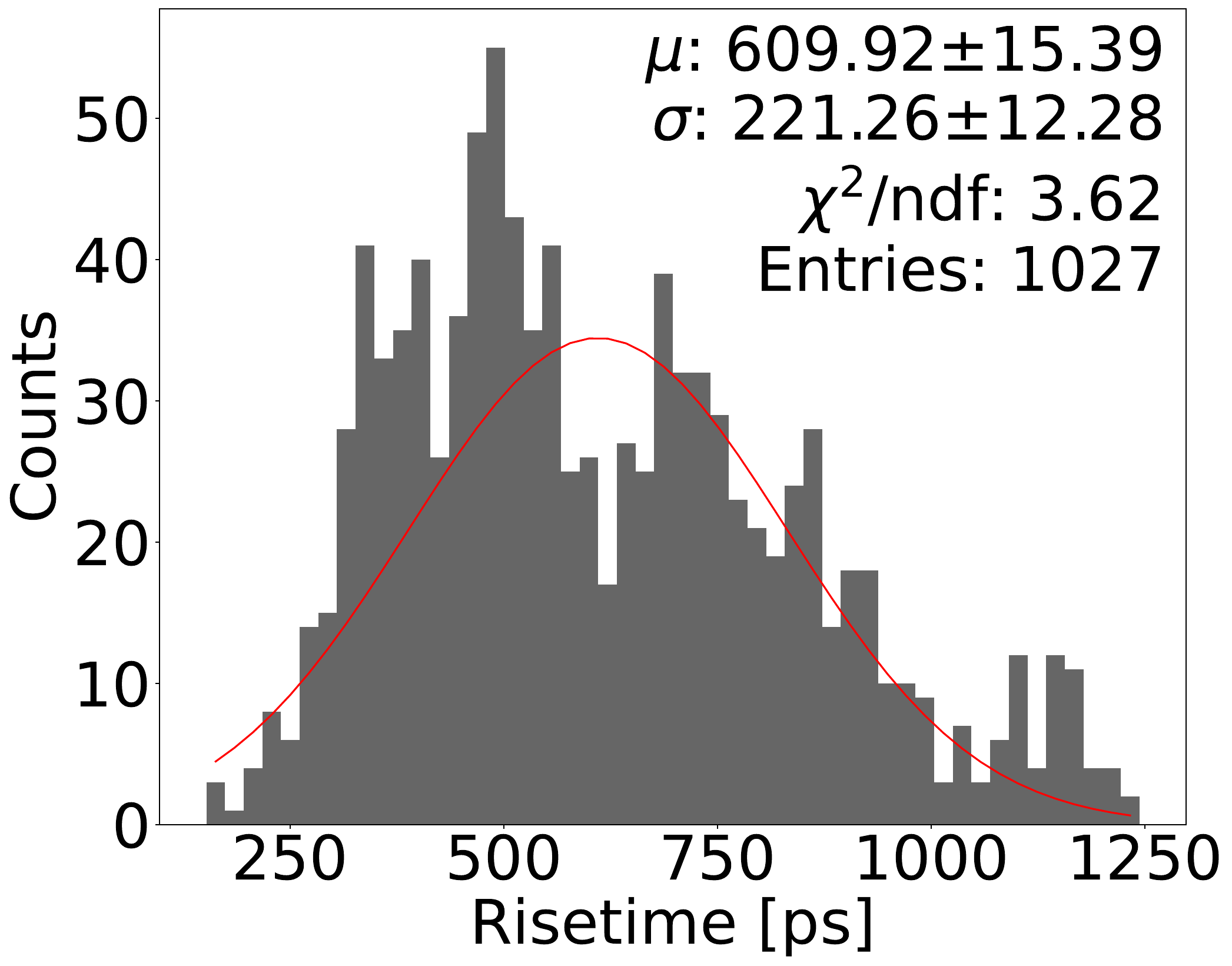}
            \caption{}
        \end{subfigure}\hspace{2em}%
        \begin{subfigure}{0.25\textwidth}
            \includegraphics[width=\textwidth]{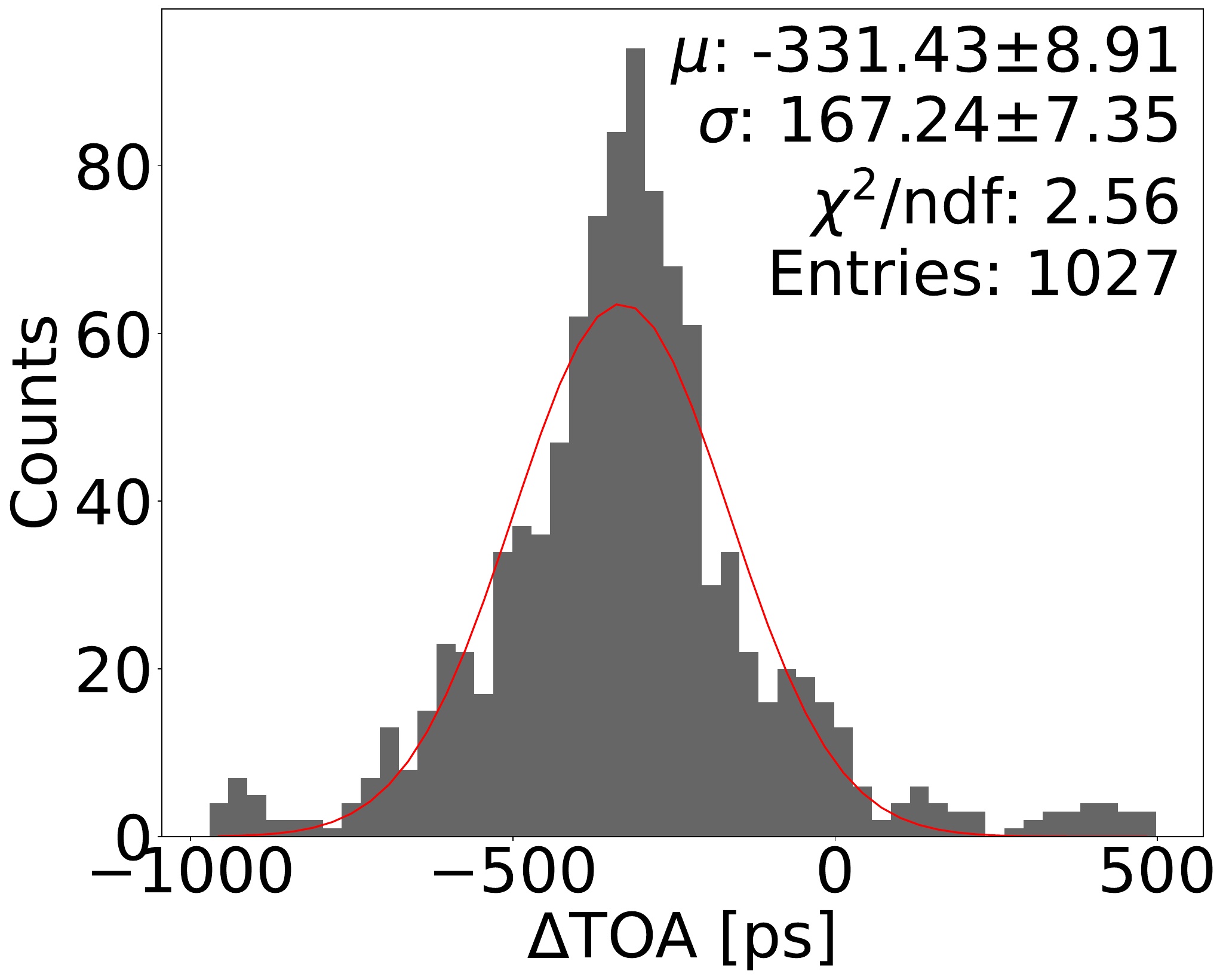}
            \caption{}
        \end{subfigure}
    \end{minipage}

	\caption{Distributions of output amplitude (a), collected charges (b), RMS noise (c), risetime (d) and ${\Delta}$TOA (e) measured with our single-channel boards and corresponding results (f-j) are measured with USTC-V1 boards.}
	\label{fig:two_rows_2}
\end{figure}

Under experimental conditions identical to those employed in the aforementioned $\beta$-source irradiation test, data were acquired over an extended accumulation period. Comprehensive analysis of the recorded dataset subsequently yielded the experimental results, which are systematically presented and discussed in the following section as shown in \autoref{fig:two_rows_2}.

Quantitative analysis of the acquired data yielded an MPV of the pulse amplitude of 64.69 mV, with a measured RMS noise level of 11.70 mV for the amplifier board. Applying the previously defined formula, the corresponding SNR was calculated as 5.53, resulting in an ENC of 0.11 fC.

Regarding time resolution, given that the reference detector exhibits a time resolution of 38.76 ps, the time resolution of our amplification board with the PIN detector is calculated to be 77.78 ps. By applying Equation~\eqref{eq:jitter}, the jitter of new preamplifier is determined to be 42.9 ps.

A detailed analysis of the USTC-V1 board coupled with the PIN detector is omitted in this work, as the measured data exhibited a poor signal-to-noise ratio that precluded reliable discrimination between the true signals and stochastic fluctuations. The results are shown here merely as a qualitative demonstration. Consequently, the USTC-V1 is considered inadequate for applications involving signal charges of less than about 1 fC.

\subsubsection{Test results with 3D silicon detector}

To validate the design, we conducted a joint test with a 3D silicon detector with a pixel size of \SI{25}{\micro\meter} and an ultra-thin active thickness of \SI{50}{\micro\meter}~\cite{ma2026thin3d}, featuring a detector capacitance of approximately \SI{15}{\femto\farad}, still employing the MCP‑PMT as the reference detector. No comparative measurement was performed in this test; we only present the results obtained from our front‑end board dedicated to the 3D silicon sensor. The experimental setup was identical to that described previously and is therefore not repeated here. All measurements were carried out in a temperature‑controlled chamber at 20 $^{\circ}$C.

\begin{figure}[h]
    \centering

    \begin{minipage}{\textwidth}
        \centering
        \begin{subfigure}{0.25\textwidth}
            \includegraphics[width=\textwidth]{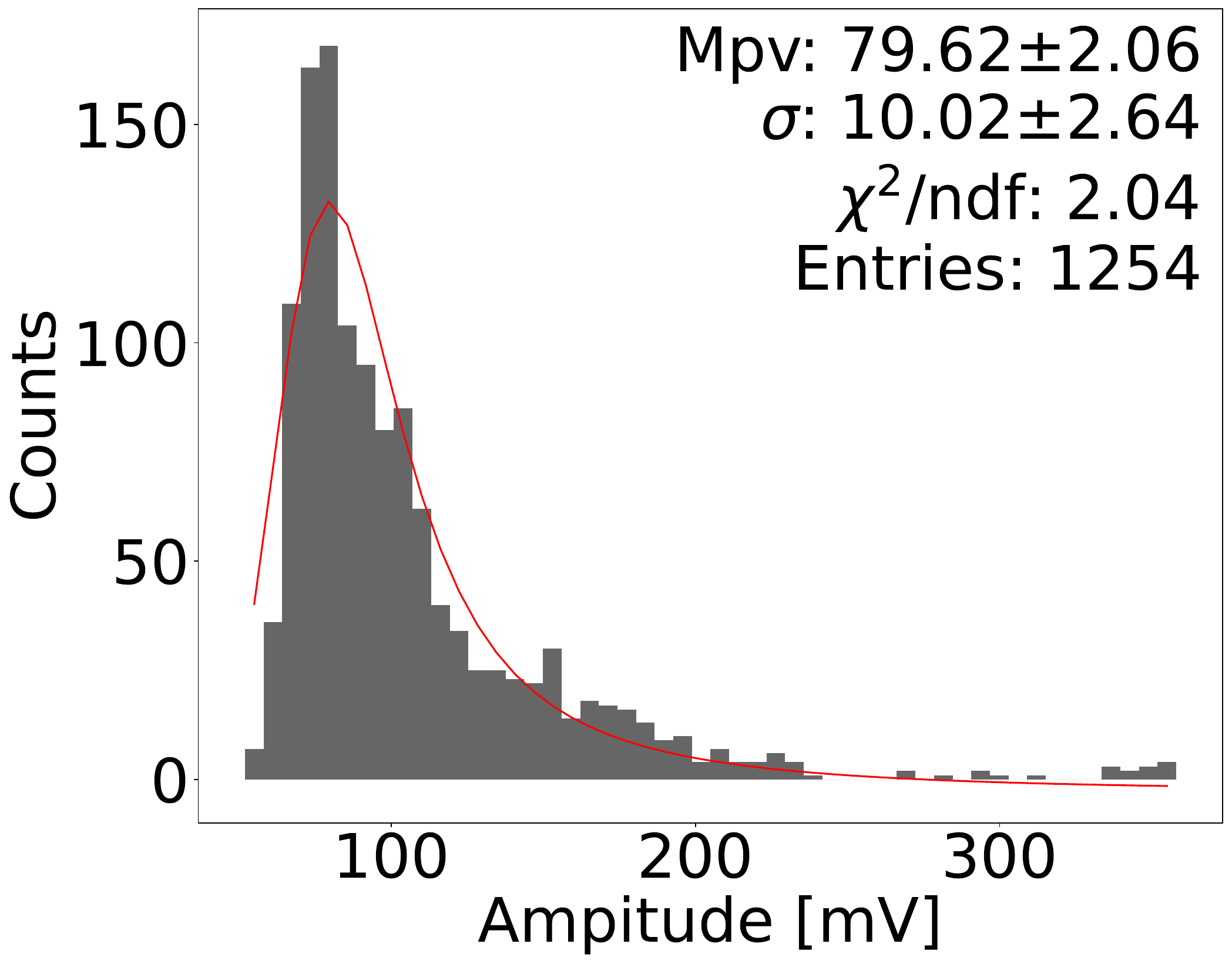}
            \caption{}
        \end{subfigure}\hspace{2em}%
        \begin{subfigure}{0.25\textwidth}
            \includegraphics[width=\textwidth]{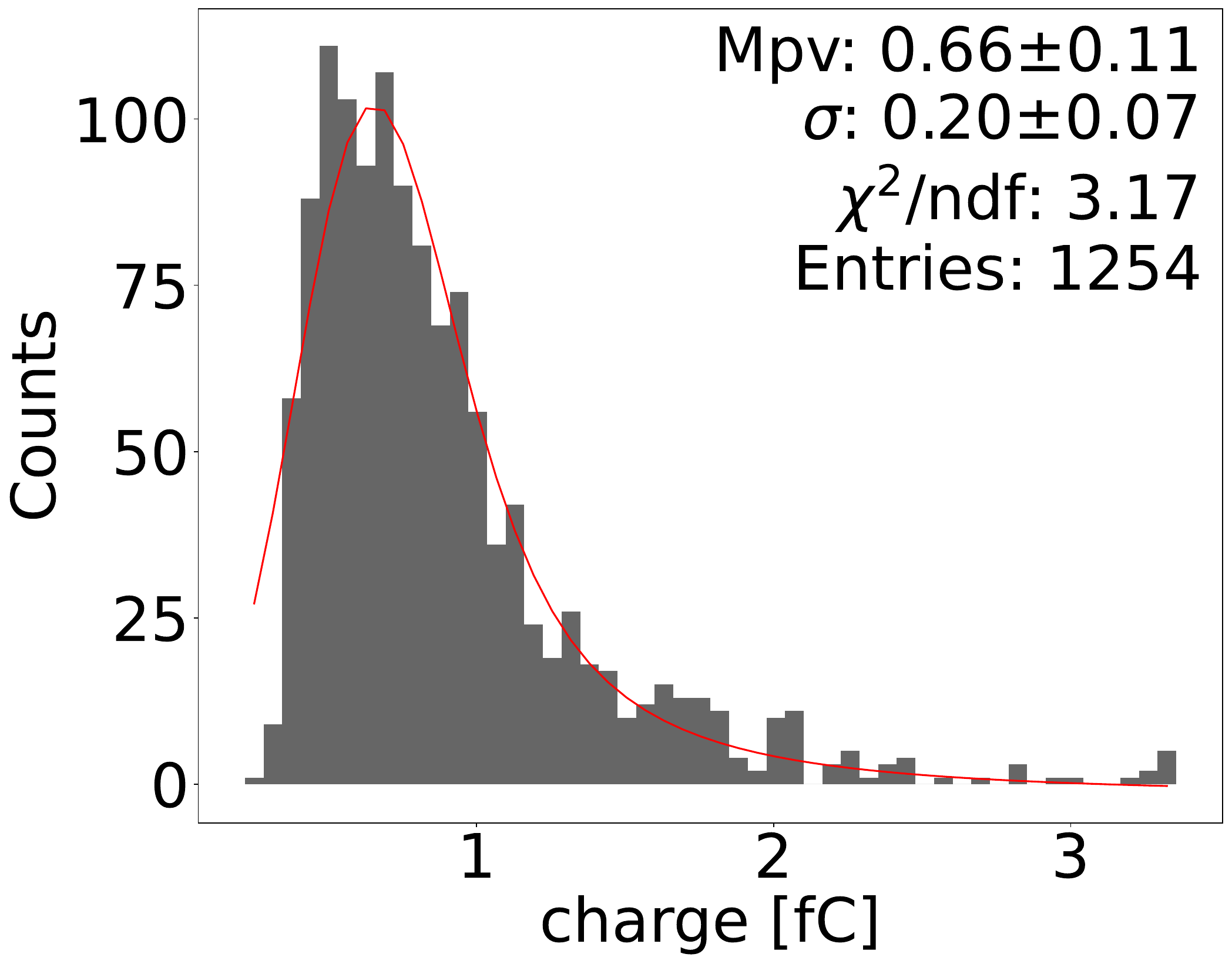}
            \caption{}
        \end{subfigure}\hspace{2em}%
        \begin{subfigure}{0.25\textwidth}
            \includegraphics[width=\textwidth]{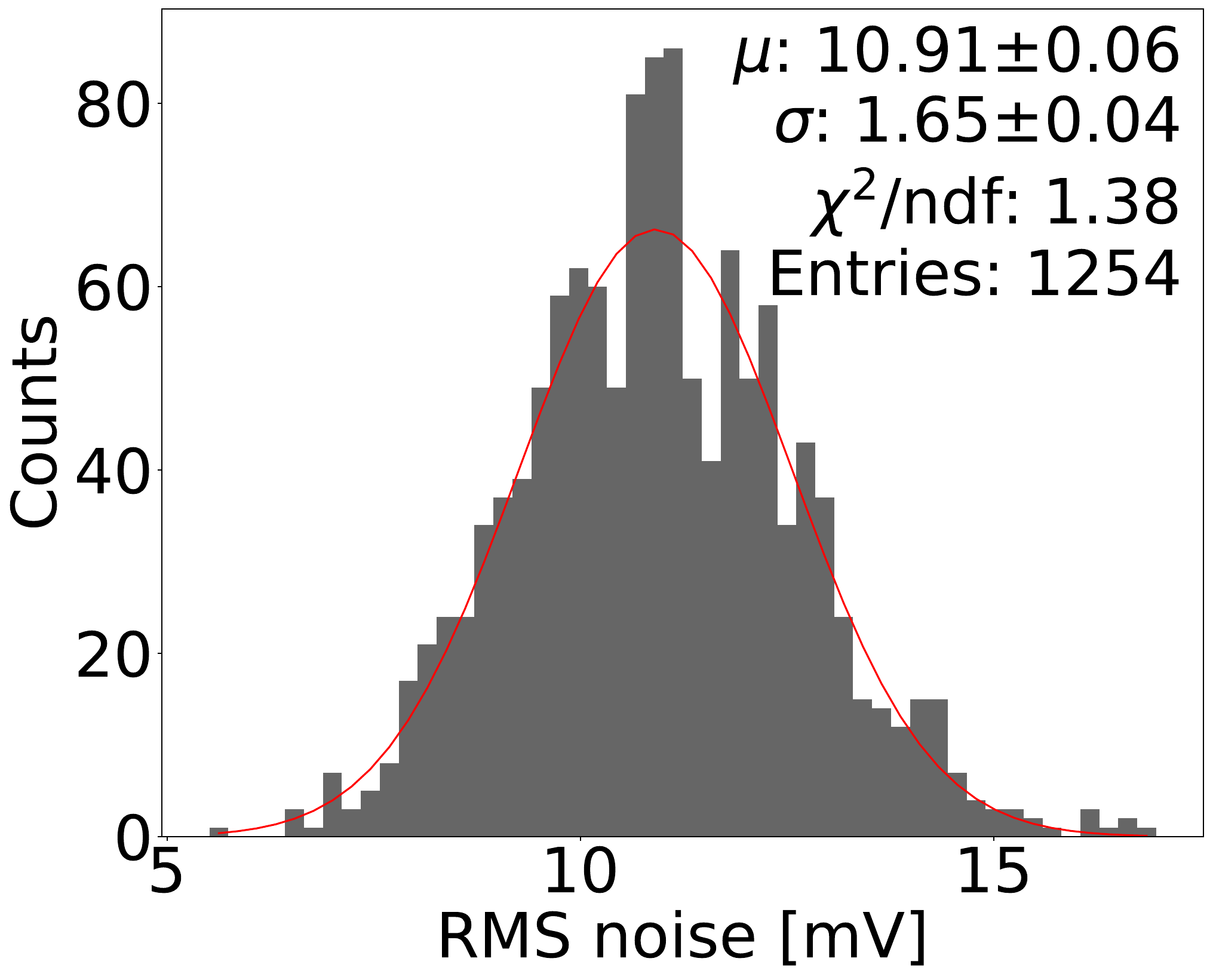}
            \caption{}
        \end{subfigure}
    \end{minipage}

    \vspace{0.5em}

    \begin{minipage}{\textwidth}
        \centering
        \begin{subfigure}{0.25\textwidth}
            \includegraphics[width=\textwidth]{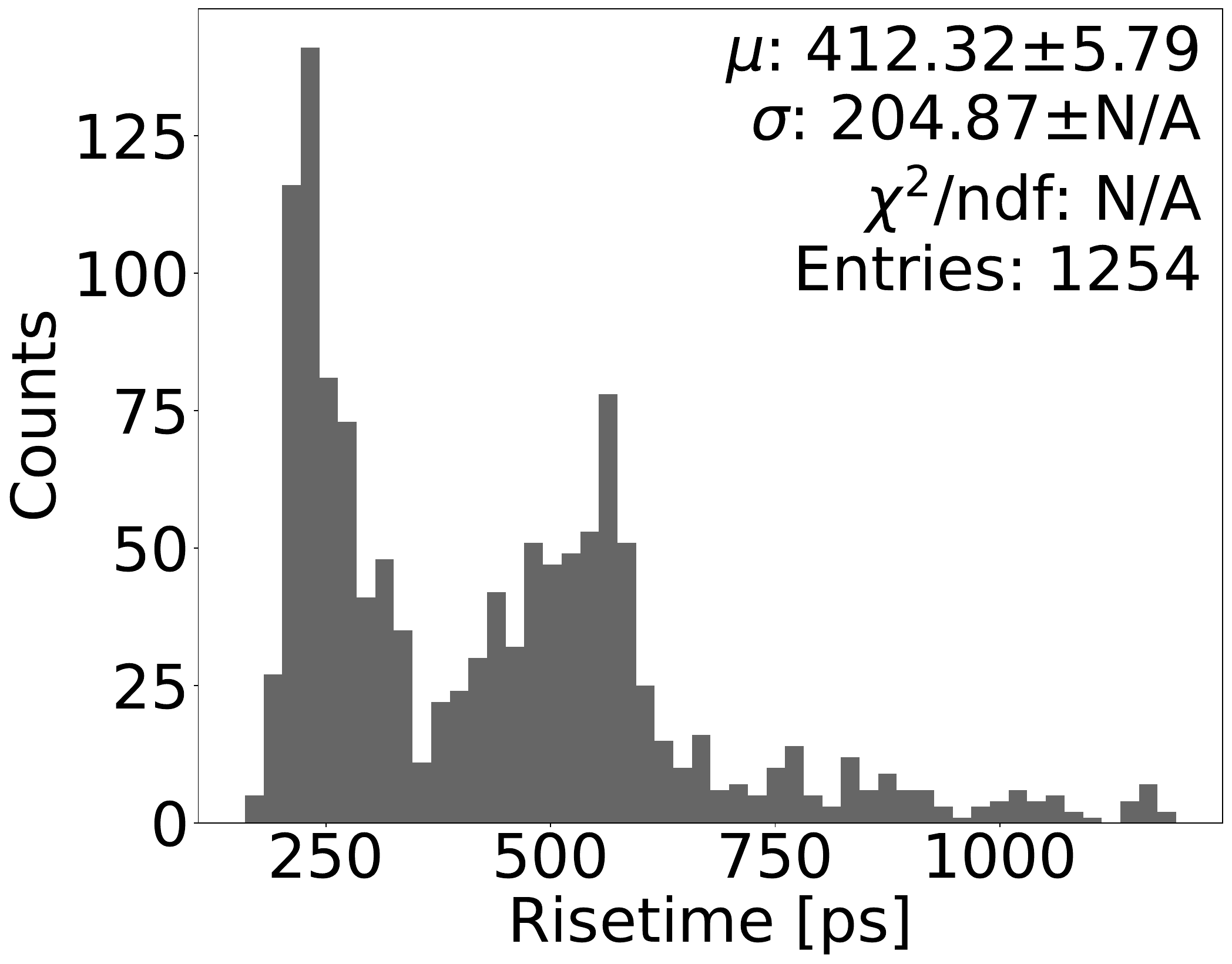}
            \caption{}
        \end{subfigure}\hspace{2em}%
        \begin{subfigure}{0.25\textwidth}
            \includegraphics[width=\textwidth]{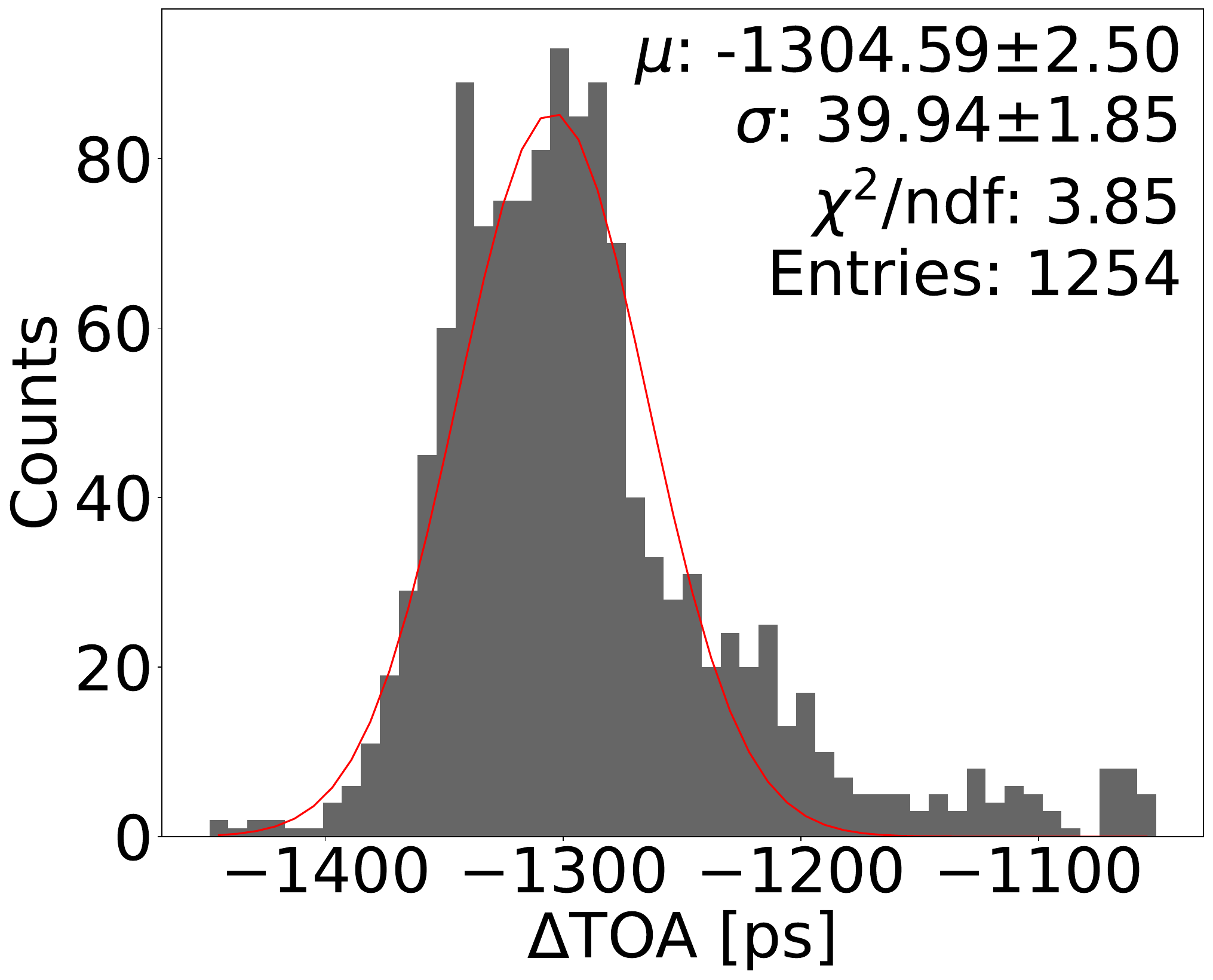}
            \caption{}
        \end{subfigure}
    \end{minipage}
    \caption{Distributions of output amplitude (a), collected charges (b), RMS noise (c), risetime (d) and ${\Delta}$TOA (e) measured with amplification board and a USTC 3D silicon detector at \qty{-70}{\volt} bias and 20 $^{\circ}$C.}
    \label{fig:3d_dist}
\end{figure}

From \autoref{fig:3d_dist}, the MPV of the amplitude was determined to be about 79.62 mV for a collected charge of 0.66 fC. The corresponding SNR was calculated as 7.30, giving an ENC of 0.09 fC. The combined time resolution, obtained from a Gaussian fit, is 39.94 ps. After deconvolving the conservative upper bound of 10 ps for the MCP-PMT reference contribution, the intrinsic timing resolution is 38.66 ps. A tail on the right side of the Gaussian peak is observed in the amplitude distribution. Similarly, the rise time distribution exhibits a distinctive bimodal structure that cannot be adequately described by a single Gaussian function. Both features are attributed to the non-uniform electric field within the sensor, which gives rise to two distinct charge-collection regimes—leading to a long tail in the amplitude distribution and two separate rise-time populations, respectively. In addition, based on the above results, the electronic jitter is estimated to be about 31.1 ps, which contributes substantially to the time resolution. The relatively large electronic jitter is primarily attributed to the limited charge output from the 3D silicon detector, which inherently restricts the achievable SNR.  Consequently, further reduction of the jitter is difficult within the present design. Substantial improvement would likely require a more advanced technology node or a detector with higher charge yield, which is beyond the scope of this work.

\subsubsection{Summary}

Table~\ref{tab:detector_comparison} summarizes the key parameters and measured performance of the three detector types.It should be noted that the LGAD and 3D silicon detector measurements both employed an MCP-PMT as the reference detector, and the reported timing resolutions are the deconvolved intrinsic values of 34.93 ps and 38.66 ps, respectively. In contrast, the PIN diode measurement used an LGAD coupled to the USTC-V1 amplifier board as the reference, and its contribution was deconvolved from the combined distribution to extract the intrinsic DUT resolution of 77.78 ps. The results confirm that the readout electronics maintain stable operation across a wide range of detector capacitances, from the ultra-low capacitance of the 3D silicon detector to the significantly larger capacitance of the PIN diode, and deliver reliable timing performance with both gain and non-gain sensors.

\begin{table}[h!]
\centering
\caption{Performance comparison of our board and USTC-V1 board}
\label{tab:detector_comparison}
\begin{tabular}{lccccc}
\toprule
Board type & \multicolumn{3}{c}{Our board} & \multicolumn{2}{c}{USTC-V1 board} \\
\cmidrule(l){2-4} \cmidrule(l){5-6}
Detector & LGAD & PIN diode & 3D silicon & LGAD & PIN diode\\
\midrule
Risetime (ps)                               &  638.13 & 432.07&  412.32 &  655.57 & 609.92 \\
Amplitude MPV (mV)                          & 929.73 & 64.68 &  79.62 & 466.00 & 29.92 \\
Collected charges MPV (fC)                  &  11.60 &  0.61 &  0.66 &  11.42 & 0.57 \\
RMS noise (mV)                         &  14.87 & 11.70 &  10.91 &  8.50 & 5.84 \\
Time resolution (ps)                        &  34.93& 77.78 &  38.66&  37.50& 167.12 \\
SNR (Amplitude MPV/ RMS noise)         &  62.52 & 5.53 &  7.30 &  54.82 & 5.12 \\
ENC (fC) (Collected charges MPV/SNR)        &   0.19 & 0.11 &   0.09 &  0.21 & 0.11 \\
\bottomrule
\end{tabular}
\end{table}

\subsection{Multi-channel preamplifier tests}

The multi-channel board shares an identical design architecture with the single-channel version; therefore, the primary objectives of its characterization were to evaluate gain uniformity across channels and to verify the absence of inter-channel crosstalk. The testing methodology employed was consistent with that used for the single-channel board, enabling the extraction of individual channel gain values, as summarized in the table below.

\begin{table}[htbp]
	\centering
	\caption{Charge gain of each channel on the multi-channel board}
	\label{tab:professional_table}
	\begin{tabular}{lcccccc}
		\toprule
		Channel               & 1  & 2  & 3  & 4  & 5  & 6  \\
		\midrule
		Charge gain(\unit{\milli\volt \cdot \nano\second \per \femto\coulomb}) & 117.79 & 116.47 & 114.54 & 115.69 & 118.92 & 118.26 \\
		\bottomrule
	\end{tabular}
\end{table}

The results demonstrate good inter-channel gain uniformity, with a mean gain of  \qty{116.95}{\milli\volt \cdot \nano\second \per \femto\coulomb} and a standard deviation of \qty{1.52}{\milli\volt  \cdot \nano\second \per \femto\coulomb}. Crosstalk is a critical concern in multi-channel designs, as any unintended signal coupling between adjacent channels would severely degrade spatial resolution. To assess this, we first injected calibration signals of varying amplitudes into the calibration port of one channel while monitoring the outputs of all other channels. No detectable signal was observed in non-injected channels, indicating the absence of electronic crosstalk through the readout chain.

Furthermore, after bonding the multi-channel board to a 5$ \times$5 LGAD detector array, we performed an optical validation test using a visible red laser. The laser spot was precisely aligned to the central pad of the detector array. Subsequent measurements showed a clear signal only in the corresponding central channel, while no measurable response was detected in the five immediately surrounding channels. Specifically, their RMS baseline fluctuations remained unchanged, and the amplitude distribution at the trigger timing was consistent with the baseline noise distribution. This observation provides strong experimental evidence that there is no significant inter-channel crosstalk—either electrical or optical—in the assembled system. Consequently, the multi-channel board is well-suited for high-fidelity position-resolved measurements.

\section{Conclusions}
A high-gain, low-noise preamplifier board has been designed specifically for low-charge  applications, and its electronic and timing performance have been thoroughly characterized through systematic testing. The preamplifier employs a TIA followed by two stages of resistive feedback gain amplification. This cascaded architecture enables a significant increase in overall voltage gain while maintaining adequate bandwidth, thereby enhancing the SNR critical for detecting weak charge signals. The TIA stage effectively converts the input current from the detector into a proportional voltage signal, while the subsequent RF gain stages provide the necessary amplification to achieve high sensitivity. Furthermore, based on this core design, a multi-channel version of the board has been developed. This multi-channel configuration provides the essential foundation for position-sensitive measurements, enabling the implementation of spatial resolution capabilities required for multi-pad detectors.

The fabricated preamplifier board has been experimentally characterized, demonstrating a charge gain of \qty{116.31}{\milli\volt  \cdot \nano\second \per \femto\coulomb} with excellent linearity over an input charge range of [0.5, 20] fC. The measured -3 dB bandwidth spans from 34.0 MHz to 594.3 MHz, indicating a broad operational frequency range suitable for various applications. Time resolution tests were conducted by coupling the preamplifier under irradiation from a \ce{^{90}Sr} radioactive source. All configurations—whether using the gain-enhanced LGAD detector or the non-gain PIN detector or 3D silicon detector—achieved favorable time resolution performance, establishing a solid foundation for future experiments.

The timing measurements with PIN and 3D detectors reveal that electronic jitter, primarily due to intrinsic noise, remains the dominant contribution when signals fall below 1 fC. This limitation is well understood and is consistent with the fundamental SNR constraints of small-charge detection. It also clearly identifies the need for either higher charge-yield sensors or more advanced fabrication technologies, thereby providing a clear roadmap for further performance enhancement.

\appendix

\acknowledgments
This work was supported by the National Natural Science Foundation of China (NSFC) under Grant No. 12535012 and the Ministry of Science and Technology of China under Grant No. 2023YFA1605901. We are grateful for all staff who participated in this work.


\bibliographystyle{JHEP}
\bibliography{biblio.bib}

\end{document}